\documentclass[aps, prc, 11pt, tightenlines, letterpaper, amsmath, amssymb, preprintnumbers, floatfix, longbibliography, nofootinbib,superscriptaddress]{revtex4-2}
\usepackage[T1]{fontenc}
\usepackage{bm}
\usepackage{xspace}
\usepackage[dvipsnames]{xcolor}
\usepackage{graphicx}
\usepackage{tabularx}
\usepackage{enumitem}
\usepackage{qcircuit}
\usepackage{braket}
\usepackage{dsfont}
\usepackage{placeins}
\usepackage[normalem]{ulem}
\usepackage{diagbox}
\usepackage{soul}
\newcolumntype{C}[1]{>{\centering\let\newline\\\arraybackslash\hspace{0pt}}m{#1}}

\usepackage{amsmath, nccmath}
\usepackage{lipsum}
\usepackage{cancel}
\usepackage{bm}

\usepackage[hypertexnames=false]{hyperref}
\hypersetup{
    colorlinks=true,       
    linkcolor=blue,          
    citecolor=blue,        
    filecolor=blue,      
    urlcolor=blue           
}

\newcolumntype{Y}{>{\centering\arraybackslash}X}

\usepackage{mathtools}

\definecolor{lavenderindigo}{rgb}{0.58, 0.34, 0.92}

\def\nab{\overrightarrow{\nabla}}

\def\llra{{\relbar\joinrel\longrightarrow}}

\def\mapup#1{{\smash{\mathop{\llra}\limits^{#1}}}}

\usepackage{stackengine}
\stackMath
\newcommand\tenq[2][1]{%
 \def\useanchorwidth{T}%
  \ifnum#1>1%
    \stackunder[0pt]{\tenq[\mumexpr#1-1\relax]{#2}}{\scriptscriptstyle\sim}%
  \else%
    \stackunder[1pt]{#2}{\scriptscriptstyle\sim}%
  \fi%
}

\def\Xint#1{\mathchoice
   {\XXint\displaystyle\textstyle{#1}}%
   {\XXint\textstyle\scriptstyle{#1}}%
   {\XXint\scriptstyle\scriptscriptstyle{#1}}%
   {\XXint\scriptscriptstyle\scriptscriptstyle{#1}}%
   \!\int}
\def\XXint#1#2#3{{\setbox0=\hbox{$#1{#2#3}{\int}$}
     \vcenter{\hbox{$#2#3$}}\kern-.5\wd0}}

\def\dashint{\Xint-}

\begin{document}
\title{Universal corrections to the superfluid gap in a cold Fermi gas}

\author{Silas R. Beane}
\email{silas@uw.edu}
\affiliation{Albert Einstein Center for Fundamental Physics, Institut f\"ur Theoretische Physik,
Universit\"at Bern, Sidlerstrasse 5, CH-3012 Bern, Switzerland}
\affiliation{Department of Physics, University of Washington, Seattle, WA 98195, USA}
\author{Zeno Capatti}
\email{zeno.capatti@unibe.ch}
\affiliation{Albert Einstein Center for Fundamental Physics, Institut f\"ur Theoretische Physik,
Universit\"at Bern, Sidlerstrasse 5, CH-3012 Bern, Switzerland}
\author{Roland C.~Farrell 
}
\email{rolanf2@uw.edu}
\affiliation{Albert Einstein Center for Fundamental Physics, Institut f\"ur Theoretische Physik,
Universit\"at Bern, Sidlerstrasse 5, CH-3012 Bern, Switzerland}
\affiliation{Department of Physics, University of Washington, Seattle, WA 98195, USA}

\preprint{NT@UW-24-09, IQuS@UW-21-083}
\date{\today}

\begin{abstract}
\noindent
A framework for computing the superfluid gap in an effective field
theory (EFT) of fermions interacting via momentum-independent contact
forces is developed.  The leading universal corrections in the EFT are
one-loop in-medium effects at the Fermi surface, and reproduce the
well-known Gor'kov-Melik-Barkhudarov result. The complete subleading
universal corrections are presented here, and include one-loop effects
away from the Fermi surface, two-loop in-medium effects, as well as
modifications to the fermion propagator.  Together, these effects are
found to reduce the gap at low densities.   Applications to neutron
superfluidity in neutron stars are also discussed.
\end{abstract}

\maketitle
\newpage{}
\pagenumbering{gobble}

\newpage{}

\section{Introduction}
\pagenumbering{arabic}
\setcounter{page}{1}
\noindent
Dense systems of cold neutrons, as found in, for example, neutron
stars and the outer shells of certain
heavy-nuclei~\cite{Dean:2002zx,Kanada-Enyo:2009bdh}, are believed to
be in a superfluid phase.  An important quantity that characterizes
this phase is the superfluid energy gap $\Delta$, that separates the
ground and first excited states in the many-body system.  Quantitative
knowledge of the gap is needed to understand properties of neutron
stars, such as their cooling rates~\cite{Page:2010aw} and spin
frequency~\cite{Watanabe:2017nzj}.  For recent reviews on the role of
superfluidity in neutrons star physics see
Refs.~\cite{Chamel:2017wwp,Haskell:2017lkl,10.1093/acprof:oso/9780198719267.003.0011,Ramanan:2020xrj}.
Despite considerable effort by nuclear theorists to compute the
superfluid gap in neutron matter, there is little consensus among
different
techniques~\cite{Ding:2016oxp,Pavlou:2016itz,Krotscheck:2021dhz,Cao:2006gq,Shen:2002pm,Schwenk:2002fq,Wambach:1992ik,Sedrakian:2003cc,Pisani:2018spx,Urban:2019ylh}.
This is likely due to a series of independent complications, including
the smallness of the gap energy relative to the scale of the strong
interaction, the absence of a clear hierarchy of scales at moderate
densities, and the intrinsic complexity of the in-medium interaction.
In parallel, there have also been ab initio approaches that determine
the gap using Quantum Monte Carlo (QMC)
simulations~\cite{Gezerlis:2007fs,Gandolfi_2022,Carlson:2012mh,Gezerlis:2009iw,Gandolfi:2008id}.
In principle, these QMC simulations do not rely on uncontrolled
approximations, and provide a powerful benchmark for nuclear theorists
to compare to.

At very-low temperatures and densities, many-body systems of fermions
are universally characterized by a momentum-independent interaction,
with a strength proportional to the free-space s-wave scattering
length $a$.\footnote{Non-universal corrections due to, for instance, effective-range or three-body interactions will be treated elsewhere.
These corrections are essential for a quantitative description of neutron matter.}  Relevant to superfluid pairing is the momentum of
fermions at the Fermi surface, $k_F$, that is related to the number
density by $\rho=N/V = k_F^3/(3\pi^2)$. In the presence of a Fermi
surface, one may then expect to develop a perturbation theory
organized in powers of the dimensionless quantity $k_F a$, while
accounting for the essential singularity at vanishing coupling due to
the BCS instability.  However, as the BCS instability is inherently
non-perturbative~\cite{PhysRev.108.1175,Polchinski:1992ed,Shankar},
the formulation of a perturbative EFT description, which by definition
is systematically improvable, is complicated even in the simplest case
of a weak finite-range
potential~\cite{Papenbrock:1998wb,Marini,Furnstahl:2006pa,Schafer:2006yf}.
In neutron matter the scattering length is very large, and the
densities where such a perturbation theory applies are not of much
physical interest.  However, in atomic physics, where the scattering
length can be tuned with Feshbach
resonances~\cite{O_Hara_2002,PhysRevLett.91.020402}, the dependence of
the gap on $k_F a$ is essential to understanding the BCS-BEC
crossover~\cite{Giorgini:2008zz,Chang:2004zza}.  In any case, the
simplicity of this universal system offers a useful theoretical
laboratory for the development of systematic methods, and will be the
focus of this work.

In BCS theory (see App.~\ref{app:scbcs} for a derivation), the superfluid gap is given in
terms of the scattering length by
\begin{equation}
\Delta_{{\rm BCS}} = \frac{8}{e^2}\omega_{k_F} \exp{ \left (\frac{\pi}{2 k_F a} \right )} \ ,
\label{eq:BCSGap}
\end{equation}
where $\omega_{k_F} \equiv k_F^2/(2M)$ is the Fermi energy with fermion mass $M$, and $e$ is
Euler's number. The effects of particle-hole screening were computed by Gor'kov-Melik-Barkhudarov (GM) in Ref.~\cite{GorkovMelik}
leading to the universal suppression,
\begin{equation}
\Delta_{{\rm GM}} \ = \ \frac{1}{(4 e)^{1/3}}\Delta_{{\rm BCS}} \ = (0.45138\ldots) \Delta_{{\rm BCS}} \ .
\label{eq:GorkovGap}
\end{equation}
In addition, further subleading corrections have been sketched out and computed for the
induced~\cite{Kohn:1965zz} p-wave gaps~\cite{Baranov,Efremov,PhysRevB.48.1097} but, as far
as the authors are aware, a full calculation of the subleading
contributions to the (s-wave) neutron superfluid gap does not exist.

A goal of this paper is to establish an EFT formulation for calculating the superfluid gap in the case of a momentum-independent potential.
In this formulation, the gap is extracted through a singularity analysis of the in-medium four-point correlation function~\cite{landau1980course}, which is determined order-by-order in perturbation theory.
Using this formulation, the subleading correction to the s-wave gap is found to be
\begin{align}
\Delta \ = \ \Delta_{{\rm GM}} \left (1\ + \ \frac{0.95238(40)}{\pi}k_F a \ + \ {\mathcal O}[(k_F a)^2] \right )\ .
\label{eq:gapfinal}
\end{align}
This result was obtained to very high accuracy by using a technique originally developed in the context of relativistic quantum field theory to numerically evaluate in-medium Feynman diagrams. See App.~\ref{app:cdNNLO} for details and references.
As $a <0$ (attractive potential), the gap is reduced relative to $\Delta_{\rm GM}$, in agreement with the QMC simulations in Ref.~\cite{Gezerlis:2007fs}.

Although the gap is inherently non-perturbative, the logarithm of the gap has a  well-defined expansion in powers of the dimensionless quantity
$\lambda = 2 k_F a/\pi$ (this quantity is referred to as the gas parameter in
Ref.~\cite{Efremov_2000}),
\begin{align}
\log{ \left ( \frac{\Delta}{\omega_{k_F}} \right )} \ &= \ \frac{c_{-1}}{\lambda} \ + \ c_{0}  \ + \ c_1 \lambda \ + \ \ldots  
\label{eq:GapPert}
\end{align}
where $c_n$ are the coefficients of the terms with $\lambda^n$.\footnote{Terms that
are exponentially suppressed in the expansion parameter $\lambda$ are neglected.}
The result in Eq.~\eqref{eq:gapfinal} gives coefficients of natural size,
\begin{align}
c_{-1} \ = \ 1 \ \ , \ \ c_0 \ = \ \frac{7}{3} \left (\log{2}\ - \ 1\right )= -0.71599...  \ \ , \ \ c_1 \ = \ 0.47619(20)\ .
\end{align}

The basis for the EFT developed in this work is a power counting scheme
for collecting the Feynman diagrams that contribute to $\log{\Delta}$
at a given order in $\lambda$.  The novel feature utilized here
is the explicit tracking of powers of $\lambda$ arising from the BCS
singularities in particle-particle loops.
By consistently counting powers of $\lambda$, it is clear that the
prefactor of the exponential in $\Delta_{{\rm BCS}}$ has no particular meaning as
it arises from an inconsistent treatment of perturbation theory: only
part of the contributions to $c_0$ are taken into account.  This has
led to confusion in the literature since the GM suppression
relative to the BCS prediction is over 50\%, and suggests that
particle-hole screening is an anomalously large effect.  
In
some sense, the GM result is the true leading order (LO) prediction for the
gap, as it is the first order where a quantitative prediction can be made.
The coefficient $c_{-1}$ only determines the
exponential dependence of the gap, and $c_0$ is needed to set the
${\cal O}(1)$ prefactor to the exponential.  
In this paper, working to LO, NLO and NNLO will correspond to calculating $c_{-1}$, $c_0$ and $c_1$ respectively.

This paper is organized as follows. Section~\ref{sec:eft} reviews the
necessary EFT ingredients. The free-space EFT which describes
low-energy fermion-fermion scattering is developed in
section~\ref{sec:fseft}. This includes the two-body scattering conventions,
as well as the scheme used to renormalize the singular interaction. The EFT
is then adapted to the in-medium calculation in
section~\ref{sec:fdeft}. In section~\ref{sec:sgpt}, the perturbative
EFT scheme is developed. Section~\ref{sec:sgptbm} and \ref{sec:sgptLO}
set up the basic methodology that motivates the power-counting scheme presented in section~\ref{sec:sgptpc}.
The NLO calculation is shown to recover the GM result
in section~\ref{sec:sgptNLO}, and the NNLO results are summarized in
section~\ref{sec:sgptNNLO}.  Finally, section~\ref{sec:conc}
provides a conclusion and a discussion of future related work.
Several clarifying derivations, and the bulk of the calculational
details are relegated to Appendices.

\section{EFT Preliminaries}
\label{sec:eft}

\subsection{Free-space EFT}
\label{sec:fseft}
\begin{figure*}
 \centerline{\includegraphics[width=0.45\textwidth]{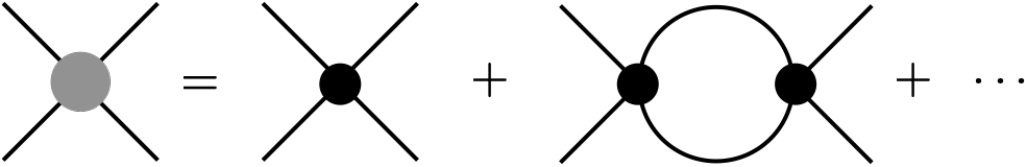}}
 \caption{Sum of Feynman diagrams contributing to fermion-fermion scattering. The black blob corresponds to the $g$ interaction from Eq.~(\ref{eq:lag}).}
\label{fig:EFTfs}
\end{figure*}

\noindent
Consider a system of spin-$1/2$ Fermions in vacuum which interact via two-body contact forces.
At very low energies, the Lagrange density takes the Galilean invariant form
\begin{eqnarray}
  {\cal L}  &=& \sum_{\sigma = \uparrow,\downarrow} \Bigg\lbrack \psi_\sigma^\dagger \left( i\hbar\partial_t + \frac{\hbar^2\nab^{\,2}}{2M}\right) \psi_\sigma - \frac{1}{2} g (\psi_\sigma^\dagger \psi_\sigma)^2 \Bigg\rbrack\, ,
    \label{eq:lag}
\end{eqnarray}
where the field $\psi_\sigma^\dagger$ creates a fermion of spin $\sigma=\uparrow,\downarrow$ and $g$ is the bare coupling constant.
In what follows, units with $\hbar =1$ are adopted. The s-wave scattering amplitude corresponding to a momentum independent interaction is
\begin{equation}
T(k)\ =\ -\frac{4 \pi}{M}
\Big\lbrack -1/a - i k \Big\rbrack ^{-1},
\label{eq:reamp}
\end{equation}
where $k=\sqrt{ME}$ is the on-shell center-of-mass (c.o.m.) momentum and $a$ is the scattering length.
Computing the scattering amplitude in the EFT from the sum of Feynman diagrams shown in Fig.~\ref{fig:EFTfs} gives
\begin{eqnarray}
T(k) &=& g+g^2\,{\mathbb{I}}(k)+g^3\,{\mathbb{I}}(k)^2+\ldots \ =\  \left( \frac{1}{g} - {\mathbb{I}}(k) \right)^{-1},
\label{eq:sla1}
\end{eqnarray}
where the geometric series has been summed and where the divergent integral
\begin{equation}
\mathbb{I}(k) \equiv  \left(\frac{\mu}{2}\right)^{3-d}  M\int \! \frac{d^d{l}}{(2 \pi)^{d}} \, \frac{1}
{{k^2}- {l^2}+i\epsilon}\ \  \mapup{\rm PDS}\ \ -\frac{M}{4\pi}\left(\mu +i k \right) \ ,
\label{eq:IEdef}
\end{equation}
has been evaluated in dimensional regularization with the power-divergence subtraction\footnote{In this scheme logarithmic divergences in $d=2$ and $d=3$
dimensions are subtracted. The limit $\mu\to 0$ then recovers the minimal subtraction ($\overline{\text{MS}}$) scheme where only the divergences in $d=3$ are subtracted.} (PDS)
scheme~\cite{Kaplan:1998tg,Kaplan:1998we}, where $d$ is the number of spatial dimensions and $\mu$ is the renormalization group
(RG) scale.  Matching the scattering amplitudes in Eqs.~(\ref{eq:reamp}) and (\ref{eq:sla1}) gives
\begin{eqnarray}
&& \frac{1}{a} = \frac{4\pi}{M g} + \mu \equiv \frac{4\pi}{M g_R}  \ ,
  \label{eq:matchsla}
\end{eqnarray}
where the renormalized coupling $g_R$ has been defined. Note that
$\lambda = (M\, k_F\, g_R)/(2\pi^2)$.  For most of the calculations in
this work it is convenient to work in the $\overline{\text{MS}}$
scheme where $\mu = 0$ and
$g_R=g$. 
However, the PDS scheme is useful as a consistency check that
the theory is renormalizable; i.e., that all the $\mu$-dependence
cancels in physical quantities.  In PDS, the bare coupling in the
Lagrangian Eq.~(\ref{eq:lag}), runs with the RG in such a way to
cancel the UV divergences ($\mu$-dependence) coming from the loop
integrals.  That is,
\begin{equation}
g(\mu) \ = \ \frac{4 \pi a}{M}\frac{1}{1-a \mu} \ = \ \frac{4 \pi a}{M}\left (1  \ + \ a \mu \ - \ {\mathcal O}\left [ (a\mu)^2\right ] \right ) \ .
\label{eq:gmu}
\end{equation}
In a consistent EFT, observables will be $\mu$-independent at each order in the expansion.
This is verified for the gap calculation to NNLO in App.~\ref{app:cdNNLO}.

\subsection{Finite density EFT}
\label{sec:fdeft}
\noindent
The zero-temperature superfluid gap is traditionally computed in the
finite-temperature Matsubara formalism (see App.~\ref{app:scbcs}) with
the zero-temperature limit taken at the end. Here, by contrast, the
zero-temperature Feynman diagram expansion will be used to compute the
in-medium four-point correlation function. That these two methods lead
to equivalent physical results is known as the Kohn-Luttinger-Ward
theorem~\cite{Kohn:1960zz,Luttinger:1960ua}, and is discussed at length
in the context of the EFT of contact forces in
Ref.~\cite{Furnstahl:2006pa,Fetter}. A consequence of working in the
zero-temperature EFT is that the chemical potential, $\mu_F$, is taken
to have its own expansion in the interaction strength with the leading
contribution given by the Fermi energy (see App.~\ref{app:mu}).
The relevant Feynman rules for computing Feynman diagrams in-medium 
in the EFT at weak coupling can be found in
Refs.~\cite{Hammer:2000xg, Furnstahl:2006pa,landau1980course}. In particular, the
interaction vertex can be taken from the Lagrange density in Eq.~(\ref{eq:lag}), and
internal lines are assigned propagators 
\begin{align}
iG_0 (k_0,{\bf k})\delta_{\alpha\gamma} &=\ i\delta_{\alpha\gamma}
    \left( \frac{\theta(k-k_F)}{k_0-\omega_{ k}+i\epsilon}
      +\frac{\theta(k_F-k)}{k_0-\omega_{ k}-i\epsilon}\right) \nonumber \\[4pt]
      &= \ \delta_{\alpha \gamma} \left (\frac{i}{k_0 - \omega_{ k} +i \delta} - 2 \pi  \delta(k_0 -\omega_{ k})\theta(k_F - k) \right ) \ ,
\end{align}
where $\alpha$ and $\gamma$ are spin indices, and $\omega_{ k} =
k^2/2M$.  The first line splits the propagator between particles and
holes and the second line splits the propagator between vacuum and
in-medium components~\cite{Kaiser:2011cg}. Arrows on fermion lines are used to
differentiate particles and holes in in-medium Feynman diagrams.

\section{Superfluid gap in perturbation theory}
        \label{sec:sgpt}
        
\subsection{Basic methodology}
        \label{sec:sgptbm}
\noindent
In a Fermi gas at zero temperature with attractive interactions, superfluid pairing is
present between particles with momentum ${\bf k}_1$ and ${\bf k}_2$
that satisfy ${\bf k}_1 = -{\bf k}_2$ and $k_1 = k_2 = k_F$.  These
momenta will be referred to as the ``BCS kinematics".  Pairing is
due to the presence of a Fermi surface, and implies that
attractive interactions between particles with BCS kinematics are
never ``weak".  This leads to the formation of Cooper pairs,
characterized by strong correlations between pairs of particles in
momentum space, and an energy gap $\Delta$ between the ground and
first excited state of the Fermi gas.

\begin{figure}
    \centering
    \includegraphics[width=0.8\textwidth]{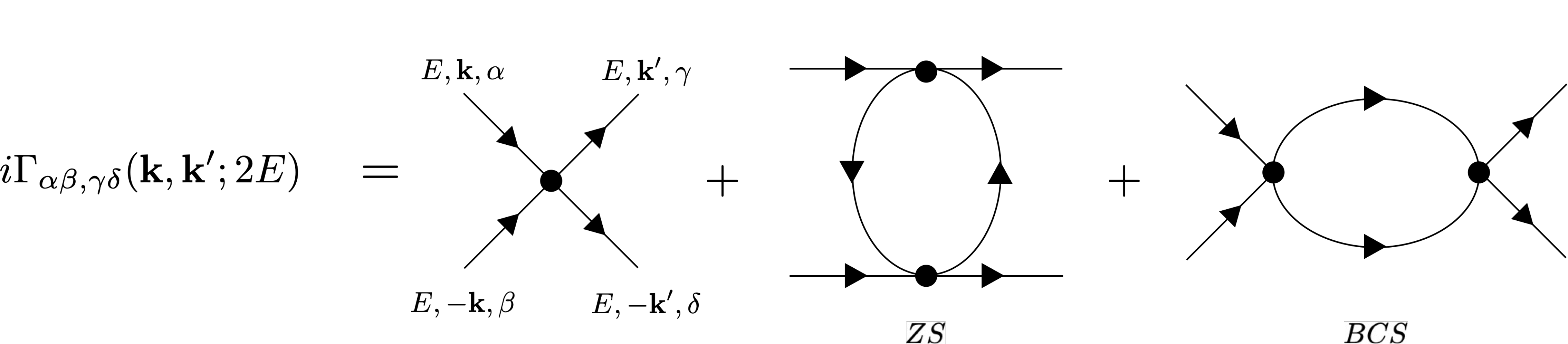}
    \caption{The connected four-point vertex function $\Gamma_{\alpha \beta, \gamma \delta}({\bf k} ,{\bf k}';2E)$ for BCS kinematics to one loop.}
    \label{fig:GamEff}
\end{figure}

In many-body perturbation theory, superfluid pairing manifests as a singularity in the 4-point vertex function $\Gamma({\bf k},{\bf
  k}';2E)$, shown to one loop in Fig.~\ref{fig:GamEff}. At one loop
order there is the Zero Sound (ZS) diagram that has
particle-hole (p-h) intermediate lines, and  the BCS diagram that
has particle-particle (p-p) intermediate lines.
The strategy for computing $\Delta$ will be to compute $\Gamma({\bf k},{\bf
  k}';2E)$ to a given order in $k_F a$, with the gap equal to the
imaginary part of the pole in total energy~\cite{landau1980course,10.1063/1.3051555,Maiti_2013} i.e. solves,
\begin{equation}
\left [\Gamma({\bf k},{\bf k}';\text{Re}[2E]+i\Delta)\right ]^{-1} \ = \ 0 \ .
\label{eq:gapPole}
\end{equation}
At a practical level, the BCS singularity
necessitates the summation of all p-p loops ---the so-called ladder
diagrams--- and therefore pairing is an inherently non-perturbative
phenomenon.  Despite this proliferation of diagrams, perturbation
theory can still be used to determine which diagrams are included 
at a given order.  The power counting used to organize perturbation
theory must account for powers of $k_F a$ coming both from the bare
interaction {\it and} from the BCS singularity.
The latter contribution
is subtle to categorize, and will be motivated with an explicit calculation of
$\Gamma$ to LO.
%

\subsection{The Gap at LO}
        \label{sec:sgptLO}
\noindent
For generic kinematics that do not lead to singularities in the BCS or ZS diagrams, the vertex function to LO is simply given by the tree diagram,
\begin{align}
\Gamma_{\alpha \beta, \gamma \delta}({\bf k} ,{\bf k}';2E) \ = \ &-g(\delta_{\alpha \gamma} \delta_{\beta \delta} \ - \ \delta_{\alpha \delta} \delta_{\beta \gamma}) \ + \ {\cal O}(g^2) \ . 
\label{eq:GamTree}
\end{align}
For BCS kinematics, it will be shown that p-p loops are no longer suppressed by powers of the coupling due to the BCS singularity.
The vertex function to one loop is given by
\begin{align}
\Gamma_{\alpha \beta, \gamma \delta}({\bf k} ,{\bf k}';2E) \ = \ &(-g+g^2 \Pi_{pp}(E))(\delta_{\alpha \gamma} \delta_{\beta \delta} \ - \ \delta_{\alpha \delta} \delta_{\beta \gamma}) \nonumber \\[4pt]
&- \ g^2\left (\Pi_{ph}({\bf k}, -{\bf k}') \delta_{\alpha \gamma} \delta_{\beta \delta}-\Pi_{ph}({\bf k}, {\bf k}') \delta_{\alpha \delta} \delta_{\beta \gamma} \right ) \ ,
\label{eq:Gam1loop}
\end{align}
where $\Pi_{pp}$ and $\Pi_{ph}$ come from evaluating the loop integrals in the BCS and ZS diagrams respectively; see Fig.~\ref{fig:GamEff}.
The $\Pi_{ph}$ terms do not contain the BCS singularity, and to this order can be dropped.
Evaluating the BCS diagram gives
\begin{align}
\Pi_{pp}(E) \ &= \ -i \int \! \frac{d^4 l}{(2\pi)^4} G_0(-l_0,-{\bf l}) \, G_0(l_0+2E,{\bf l}) \nonumber \\[4pt]
& = \ -M\int \! \frac{d^{3} {\bf l}}{(2\pi)^{3}} \frac{1}{2ME - l^2 + i \epsilon}  \ + \ 2 M \dashint \frac{d^3 {\bf l}}{(2\pi)^3} \frac{\theta(k_F - l)}{2ME - l^2} \ .
\label{eq:piPP}
\end{align}

The first term in the second line of Eq.~(\ref{eq:piPP}) is the same as in vacuum, and is given by Eq.~(\ref{eq:IEdef}).
The second term in the second line of Eq.~(\ref{eq:piPP}) has a
logarithmic divergence at $2E = 2\omega_{k_F}$, that is regulated by
the imaginary piece of the energy $(i \Delta)$.  This singularity can
be extracted through an integration by parts~\cite{Fetter},
\begin{equation}
2M\dashint \frac{d^3 {\bf l}}{(2\pi)^3} \frac{\theta(k_F - l)}{2ME - l^2} \ = \ \frac{M}{2 \pi^2} \left [ \dashint^{k_F}_0 dl \, \log{\left (2ME-l^2 \right)} \ - \ k_F\log{(2ME-k_F^2)}  \right ]\ .
\label{eq:c0IBP}
\end{equation}
The first term is finite for $E=\omega_{k_F}$, and a perturbatively
small $\Delta$ can be expanded in a power series whose coefficients
can depend on $\log \Delta$.  This is not the case for the second
term, which gives $\log \left (i M \Delta \right)$.  Looking ahead,
the solution for the gap in Eq.~(\ref{eq:BCSGap}) reveals that while
powers of $\Delta$ are exponentially small in $g$, $\log{\Delta}$ has
an expansion in $g$ that starts at ${\cal O}(g^{-1})$.  This implies
that one can set $E=\omega_{k_F}$ everywhere except in terms of the
form $\log{(E-\omega_{k_F})}$.  Combined with the additional factor of
$g$ from the vertex, this piece of the p-p loop integral can formally
be taken to be of the same order as the tree level contribution.  In
fact, any number of p-p loops enter at the same order, and LO consists
of the sum of an infinite number of diagrams.  A key observation, that
will simplify higher order calculations, is that the $\log{\Delta}$
piece of this loop integral occurs when the loop momenta are on-shell
and in the vicinity of $l=k_F$, since it arises from the boundary of the
integral.\footnote{That the internal lines are on-shell is easiest to
  see if the c.o.m. energy in the loop is split symmetrically as $l_0
  +E$ and $-l_0 +E$.  In this case, $l_0 =0$ when $l=k_F$ and the
  energy of each internal line is $E=\omega_{k_F}$.}

The p-p loops (ladder diagrams) do not mix partial waves or spin projections, and can be summed as a geometric series,
\begin{equation}
\Gamma_{\alpha \beta, \gamma \delta}({\bf k} ,{\bf k}';2E) \ = \ -\frac{g}{1+g \Pi_{pp}(E)}
(\delta_{\alpha \gamma} \delta_{\beta \delta} \ - \ \delta_{\alpha \delta} \delta_{\beta \gamma}) \ + \ {\mathcal O}(g^2)\ .
\label{eq:LadderSum}
\end{equation} 
For future convenience, the whole $\Pi_{pp}(E)$ has been resummed, but, to this order, only the $\log{\Delta}$ piece should be kept.
Solving Eq.~(\ref{eq:gapPole}) gives
\begin{equation}
0 = 1- \ g \frac{M k_F}{2\pi^2}\log{ \Delta} + {\mathcal O}(g) \ ,
\label{eq:LOgapEQN}
\end{equation}
with solution
\begin{equation}
\Delta_{{\rm LO}} \sim \exp{\left ( \frac{\pi}{2 k_F a} \right )} \ ,
\label{eq:C0gapLO}
\end{equation}
where $g=4\pi a/M$ has been used.
Crucially, the prefactor of the exponential \textit{cannot be
  determined} at this order.  The exponential dependence agrees with
Eq.~(\ref{eq:GorkovGap}), and the resummation of the p-p loops
predicated on $\log{\Delta}$ effects beginning at ${\mathcal O}(g^{-1})$ is
consistent.

\subsection{Power Counting}
      \label{sec:sgptpc}
\begin{figure}
    \centering
    \includegraphics[width=0.8\textwidth]{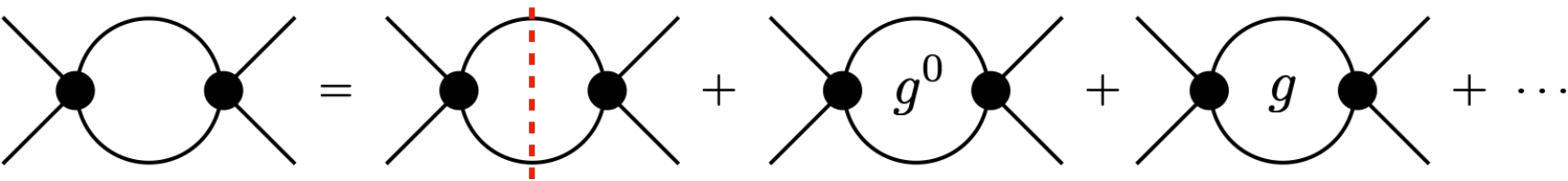}
    \caption{Diagrammatic expansion of the p-p loops in powers of $g$. The red dashed line represents the ${\cal O}(g^{-1})$ contribution that has loop four-momentum at the Fermi surface and on-shell.}
    \label{fig:pp_bubble}
\end{figure}
\begin{figure}
    \centering
    \includegraphics[width=0.8\textwidth]{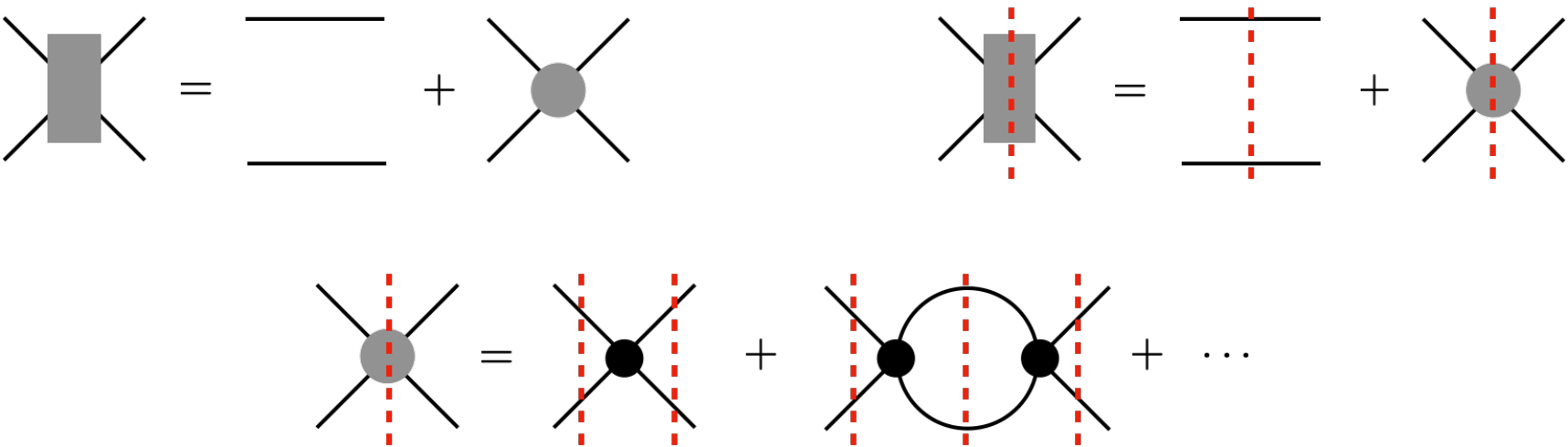}
    \caption{Resummed classes of diagrams. The grey blob is defined in Fig.~\ref{fig:EFTfs}, and a dashed red line corresponds to evaluating the kinematics on-shell and at the Fermi surface.
    A dashed line running through a p-p loop represents the leading ${\mathcal O}(g^{-1})$ contribution.
    }
    \label{fig:diagramDefined}
\end{figure}

\noindent
The LO calculation presented in the previous section demonstrated that in a consistent power-counting scheme, p-p
loops are assigned powers of $g$.  To make the power counting manifest, it
is beneficial to expand $\Pi_{pp}(E)$ as,
\begin{align}
\Pi_{pp}(E) \ = \ \Pi^{(g^{-1})}_{pp} \ + \ \Pi^{(g^{0})}_{pp} \ + \ \Pi^{(g)}_{pp} \ + \ldots \ ,
\end{align}
where the $E$ dependence on the RHS has been dropped for simplicity.  
This is shown diagrammatically in Fig.~\ref{fig:pp_bubble}.
With this
identification, it is possible to collect the Feynman diagrams that
contribute to the vertex function $\Gamma$ at a given order in $g$.  A
key feature is that any sub-diagram connected to $\Pi^{(g^{-1})}_{pp}$
can be evaluated on-shell and at the Fermi surface, since these are the
kinematics that give rise to $\log{\Delta}$.  This will simplify the
NLO and NNLO calculations presented below.  The diagrams that
contribute to LO and NLO are given in Fig.~\ref{fig:LONLOGap}, with sub-diagrams defined in Fig.~\ref{fig:diagramDefined}.\footnote{Note that the grey box with a dashed red line defined in Fig.~\ref{fig:diagramDefined} is counted as ${\cal O}(g^0)$ on external lines but ${\cal O}(g^{-1})$ on internal lines.}
For the NNLO calculation it is convenient to treat LO, NLO
and NNLO together as shown in Fig.~\ref{fig:NNLOGap}. A breakdown of the NNLO diagrams is given in Figs.~\ref{fig:NNLOdiag} and \ref{fig:NNLOV}.

\subsection{The Gap at NLO}
      \label{sec:sgptNLO}
\begin{figure}
    \centering
    \includegraphics[width=0.8\textwidth]{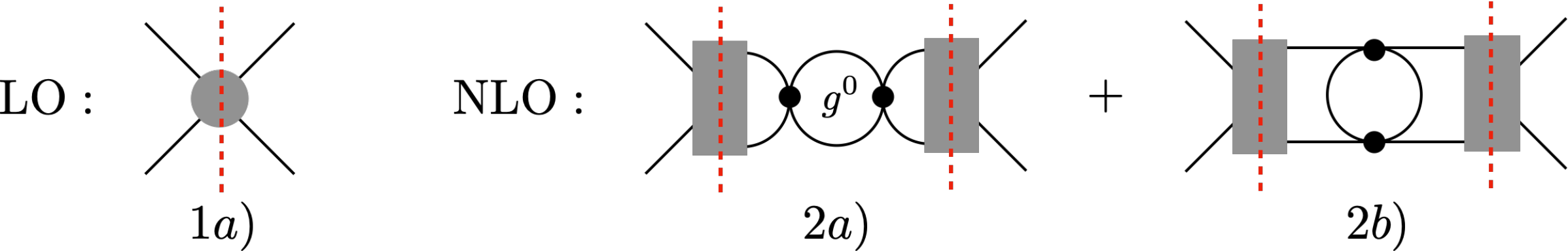}
    \caption{The diagrams that contribute to the vertex function for BCS kinematics at LO and NLO.}
    \label{fig:LONLOGap}
\end{figure}
\noindent
At NLO there are two new diagrams contributing to the vertex function, as shown in Fig.~\ref{fig:LONLOGap}.
Diagram 2a) includes a single insertion of $\Pi^{(g^{0})}_{pp}$, and summing both LO and NLO gives,
\begin{align}
\Gamma_{1a} + \Gamma_{2a} \ = \ -\frac{g}{1+g \Pi_{pp}^{(g^{-1})}} \left (1- \frac{g \Pi_{pp}^{(g^0)}}{1+g \Pi_{pp}^{(g^{-1})}}\right ) \ .
\label{eq:Gam1a2a}
\end{align} 
Diagram 2b) involves an infinite sum of p-p loops, followed by a p-h (ZS) bubble, followed by another infinite sum of p-p loops.
At this order, only $\Pi_{pp}^{(g^{-1})}$ contributes to each p-p loop, and therefore the p-h bubble can be evaluated with external legs on-shell and at the Fermi surface.
This results in,
\begin{align}
\Gamma_{2b} \ = \ \frac{-V^{(g^2)}}{\left ( 1+g\Pi_{pp}^{(g^{-1})}\right )^2}  \ ,
\label{eq:Gam2b}
\end{align}
where $V^{(g^n)}$ is the s-wave component of the two-particle irreducible (2PI) potential that scales as $g^n$, and is evaluated for external legs on-shell and at the Fermi surface.

The full 2PI potential is determined from the sum of all 2PI diagrams and contains projections onto all partial waves.
The ${\mathcal O}(g)$ contribution comes from the tree diagram, and the ${\mathcal O}(g^2)$ contribution comes from the ZS diagram in Fig.~\ref{fig:GamEff}.
Evaluating these diagrams in the c.o.m. frame one finds (note the opposite sign relative to $\Gamma$),
\begin{equation}
V^{(g^2)}_{\alpha \beta, \gamma \delta}({\bf k}, {\bf k}';\omega) \ = \  g^2\left (\Pi_{ph}({\bf k}, -{\bf k}';\omega) \delta_{\alpha \gamma} \delta_{\beta \delta}-\Pi_{ph}({\bf k}, {\bf k}';\omega) \delta_{\alpha \delta} \delta_{\beta \gamma} \right ) \ ,
\label{eq:NLO2PIFull}
\end{equation}
where
\begin{align}
    \Pi_{ph}({\bf k}, {\bf k}';
    \omega) \ &= \ i \int \! \frac{d^4 l}{(2\pi)^4} \left [ G_0(l_0,{\bf l}) \, G_0(l_0+\omega,{\bf l}+{\bf k}-{\bf k}')\right ] \nonumber \\[4pt]
    & \stackrel{\omega=0}{=} \ - 2 M \int \! \frac{d^3 l}{(2\pi)^3}  \frac{\theta(k_F - l)}{({\bf l}-{\bf k})\cdot({\bf k } - {\bf k}')}\ ,
    \label{eq:Piph}
\end{align}
with $\omega = k_0 - k_0'$ the energy transfer.
Relevant for $V^{(g^2)}$ are kinematics with $k=k'=k_F$ and $\omega=0$ giving,
\begin{equation}
\Pi_{ph}(q)  \ = \ \frac{M k_F}{4\pi^2} \left ( 1 \ + \ \frac{4-q^2}{4q} \log{ \left \lvert \frac{2+q}{2-q} \right \rvert}\right )\ ,
\label{eq:statLind}
\end{equation}
where $q = \lvert {\bf k} - {\bf k}' \rvert /k_F =
\sqrt{2-2\cos{\theta}}$ and $\theta$ is the angle between ${\bf k}$
and ${\bf k}'$.\footnote{ This depends non-analytically on
  $\cos{\theta}$, and therefore possesses all partial waves.  This is
  true even for the momentum-independent s-wave potential used in this
  work, and the attractive potential induced in higher partial waves
  leads to the Kohn-Luttinger effect~\cite{Kohn:1965zz} for $a>0$.}
The potential can now be expanded onto partial waves as
\begin{equation}
V_{\alpha \beta,\gamma \delta}(\theta) = \begin{cases}
\sum_l (2l+1)P_l(\theta) V_{l} (\delta_{\alpha \gamma}\delta_{\beta \delta} \ - \ \delta_{\alpha \delta} \delta_{\beta \gamma}) & l \ {\rm even} \ ,  \\[5pt]
\sum_l (2l+1)P_l(\theta) V_{l} (\delta_{\alpha \gamma}\delta_{\beta \delta} \ + \ \delta_{\alpha \delta} \delta_{\beta \gamma}) & l \ {\rm odd} \ ,
\end{cases}
\label{eq:partialWave}
\end{equation}
where the $V_l$ are obtained by integrating against the relevant Legendre polynomial $P_l(\theta)$.
Projecting onto the s-wave results in~\cite{Maiti_2013,Baranov} 
\begin{align}
V^{(g^2)} \ &= \frac{g^2}{2} \int_0^2 \! dq \, q \Pi_{ph}(q) \ = \frac{M g^2 k_F}{8 \pi^2} \frac{4}{3}\left (1 \ + \  2\log{2} \right )  \ . 
\label{eq:l0Project}
\end{align}

Adding together Eq.~(\ref{eq:Gam1a2a}) and Eq.~(\ref{eq:Gam2b}) gives the NLO gap equation,
\begin{align}
0 &= \left [\Gamma_{1a} + \Gamma_{2a} + \Gamma_{2b}\right ]^{-1} \nonumber \\
& = 1 + g\left (\Pi_{pp}^{(g^{-1})} + \Pi_{pp}^{(g^{0})} \right ) - \frac{V^{(g^2)}}{g} \ + \ {\mathcal O}(g^2) \nonumber \\
&= 1 +\lambda \left [ \log{\left (\frac{8 \omega_{k_F}}{e^2 \Delta} \right )} - \frac{1}{3}(1+2\log{2})\right ] \ ,
\end{align}
where the quantity
\begin{equation}
    g \left (\Pi_{pp}^{(g^{-1})} + \Pi_{pp}^{(g^{0})}\right ) \ = \ \lambda \log{\left (\frac{8 \omega_{k_F}}{e^2 \Delta} \right )} \ ,
    \label{eq:gPipp}
\end{equation}
has been determined from the integral in Eq.~(\ref{eq:piPP}) using the $\overline{\text{MS}}$ scheme.
Solving for the gap recovers the GM result~\cite{GorkovMelik},
\begin{equation}
    \Delta_{\text{NLO}} \ = \ \Delta_{\text{GM}} \ = \ \frac{8 \omega_{k_F} }{e^2 (4 e)^{1/3}} \exp{\left ( \frac{\pi}{2 k_F a} \right )}\ .
\end{equation}
%

\subsection{The Gap at NNLO}
      \label{sec:sgptNNLO}
\begin{figure}
    \centering
    \includegraphics[width=\textwidth]{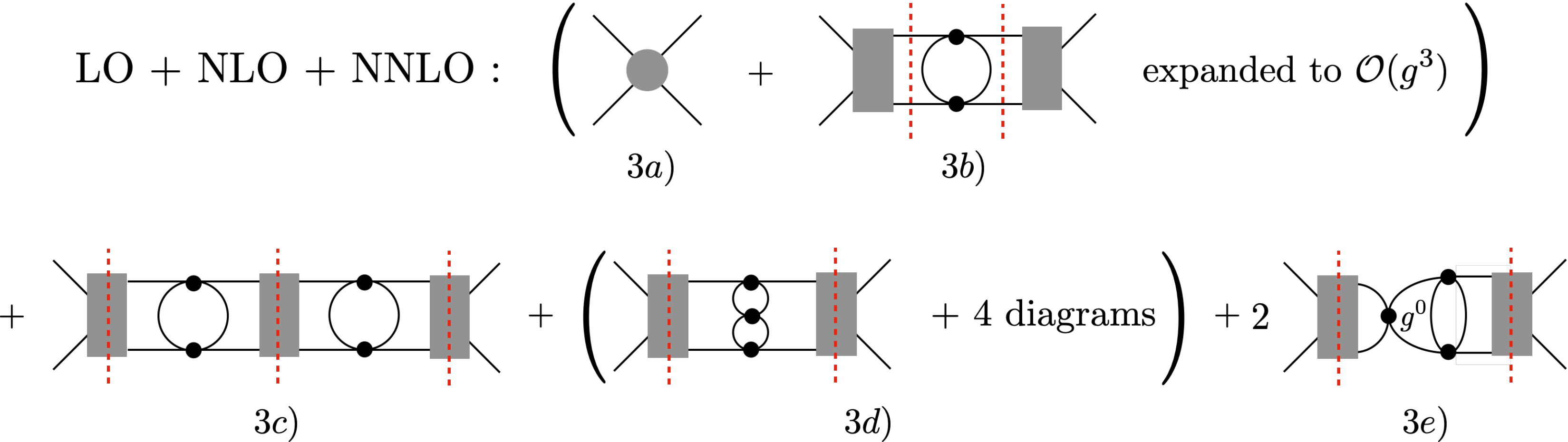}
    \caption{The diagrams that contribute to the vertex function for BCS kinematics through NNLO. 
    The ``+ 4 diagrams" corresponds to the 4 other diagrams contributing to the NNLO two-particle-irreducible potential shown in Fig.~\ref{fig:NNLOV}.
    At this order there are also corrections to the propagator, which affect 3a), and are not explicitly shown.}
    \label{fig:NNLOGap}
\end{figure}
\noindent
For the NNLO calculation it is convenient to treat LO, NLO and NNLO together. There are five types of diagrams that contribute to the vertex function, as shown in Fig.~\ref{fig:NNLOGap}.
The evaluation of the diagrams $\Gamma_{3a,b,c,d,e}$, is treated in App.~\ref{app:cdNNLO}, and the NNLO gap equation is
\begin{align}
&0 \ = \ \left (\Gamma_{3a} + \Gamma_{3b}+\Gamma_{3c}+\Gamma_{3d}+\Gamma_{3e} \right )^{-1} \nonumber \\
& = \ 1 + g\left (\Pi_{pp}^{(g^{-1})} + \Pi_{pp}^{(g^{0})}+\Pi_{pp}^{(g)} \right) - \frac{1}{g}\left ( V^{(g^2)} + V^{(g^3)}\right ) +  2{\cal I}\left [\Pi^{(g^0)}_{pp}\,V^{(g^2)}\right ] \nonumber \\
& = \ 1 +\lambda \left [ \log{\left (\frac{8 \omega_{k_F}}{e^2 \Delta} \right )} - \frac{1}{3}(1+2\log{2})\right ] \ - \ 2.231993(16) \, \lambda^2 \ + \ 1.83549(20) \, \lambda^2 \ ,
\label{eq:NNLOGapEqn}
\end{align}
where ${\cal I}\left [\Pi^{(g^0)}_{pp}\,V^{(g^2)}\right ]$ accounts
for the effects of $V^{(g^2)}$ evaluated for kinematics away from the
Fermi surface. The uncertainties are due to Monte Carlo integration.

At this order, it is also necessary to consider modifications to the
propagator, i.e. the self energy.\footnote{The effects of the chemical
  potential shifted away from $\omega_{k_F}$ do not change the NNLO
  calculation as shown in App.~\ref{app:mu}.}  These effects can be
parameterized by an effective mass $M^{\star}$ and wavefunction
renormalization $Z$ given by~\cite{landau1980course}
\begin{equation}
    \frac{M^{\star}}{M} \ = \ 1+\lambda^2\frac{2}{15}(7\log{2}-1) \ + \ \mathcal{O}(\lambda^3)  \ \ , \ \ Z \ = \ 1-\lambda^2\log{2}\ + \ \mathcal{O}(\lambda^3) \ .
\end{equation}
Taking this into account, and solving for the gap, gives
\begin{align}
\Delta_{\text{NNLO}} \ &= \ \frac{8 \omega_{k_F}}{e^2(4e)^{1/3}} (1-0.39650(20)\lambda)\exp{\left (\frac{\pi}{2 k_F a}\frac{M}{M^{\star}}\frac{1}{Z^2}\right )} \nonumber \\
& = \frac{8 \omega_{k_F}}{e^2(4e)^{1/3}}(1+0.47619(20)\lambda) \exp{\left (\frac{\pi}{2 k_F a}\right )} \nonumber \\
& = \Delta_{\text{NLO}}\left (1+\frac{0.95238(40)}{\pi}k_F a \right ) \ .
\label{eq:gapNNLO}
\end{align}
For an attractive interaction ($a<0$) the gap at NNLO is reduced relative to NLO. 
In neutron-neutron scattering, the scattering length is $a^{-1}\approx-10.7 \, \text{MeV}$~\cite{deSwart:1995ui}, and perturbation theory is expected to break down around $k_F \approx 17 \, \text{MeV}$.

\begin{figure}
    \centering
    \includegraphics[width=0.75\textwidth]{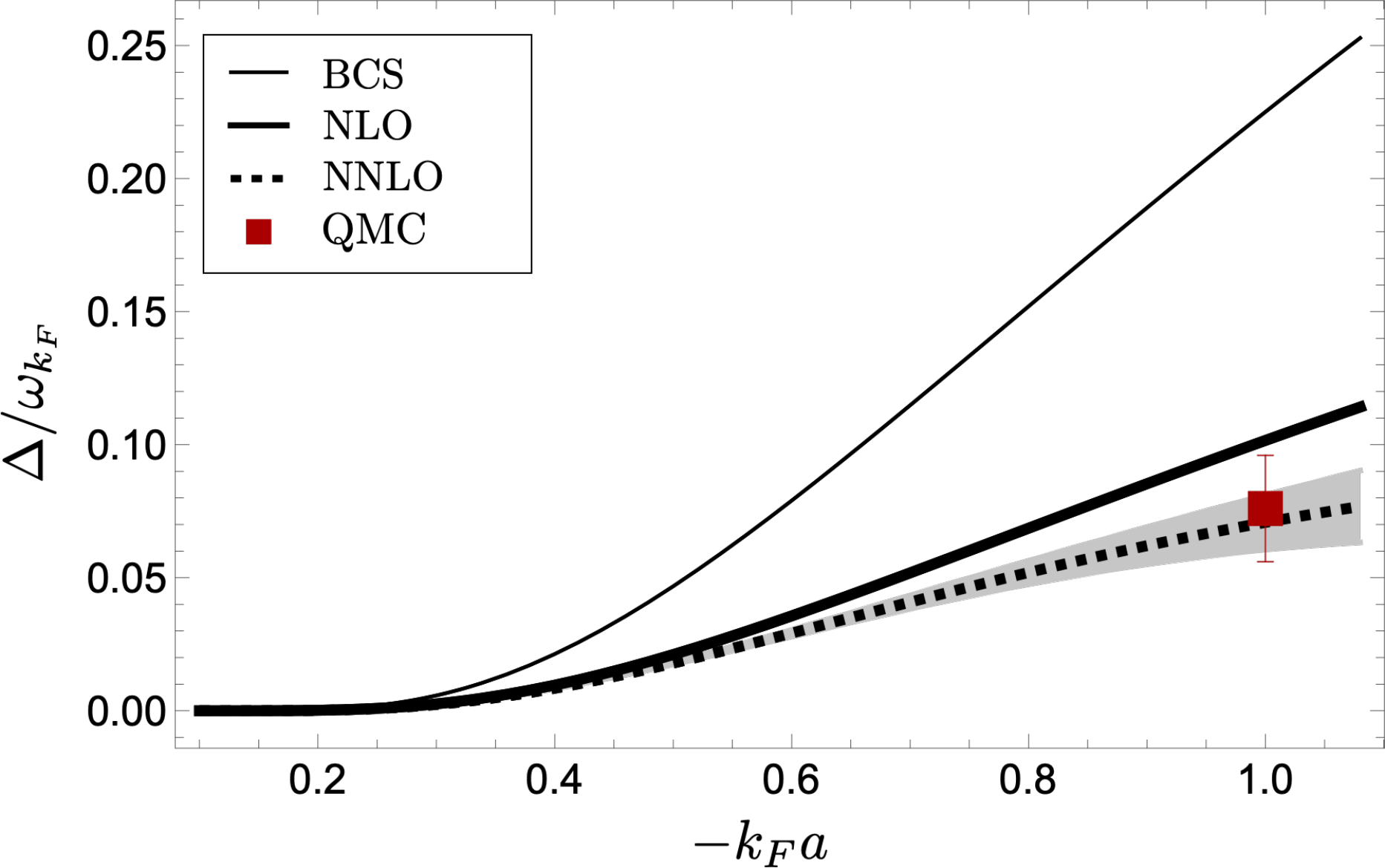}
    \caption{The BCS, NLO and NNLO superfluid gaps compared to the QMC prediction of Ref.~\cite{Gezerlis:2007fs}. 
    The error band on the NNLO calculation is estimated assuming a N$^3$LO correction with a coefficient of magnitude 1.}
    \label{fig:gapCompare}
\end{figure}

This result can be compared to Ref.~\cite{Gezerlis:2007fs} where the
gap is determined with QMC.  The QMC results have one point at $k_F a
= -1$ that is expected to be within the perturbative regime, but may
be marginal.  The gap at NLO, NNLO, as well as the BCS prediction are
compared to QMC in Fig.~\ref{fig:gapCompare}.  The error band in the
NNLO gap prediction comes from assuming that the coefficient of the
N$^3$LO correction (i.e. the next $\mathcal{O}(k_F a/\pi)^2$ term in
Eq.~\ref{eq:gapNNLO}) has magnitude 1. The QMC result lies closer to
the central value of the NNLO prediction, but the large error bars on
both the theory and QMC prediction make it difficult to make any
definite conclusions.  And, of course, the theory error bars may not be
sufficiently conservative. More precise QMC simulations across a range
of $k_F a$, and preferably at smaller values, will be needed to
confirm the validity of the perturbation theory presented here.
It is noteworthy that recent work~\cite{Pisani:2018spx}, which incorporates
effects beyond the GM approximation, and extends from weak coupling
through the BCS/BEC crossover region, is also in good agreement
with the QMC result at $k_F a = -1$.

%

\section{Conclusion}
\label{sec:conc}

\noindent 
The superfluid (or superconducting) gap energy is a hallmark feature
of fermions with attractive interactions at finite densities and low
temperatures.  This work has established an EFT framework for
computing the superfluid energy gap for fermions with momentum
independent interactions.  The momentum independent interaction
captures the universal low-energy behavior of fermions with finite
range forces, and has applications across a wide range of scales,
e.g. to many-body systems of electrons, atoms and/or nucleons.  A NNLO
calculation in the EFT revealed new, universal corrections, to the
GM~\cite{GorkovMelik} gap prediction.  These corrections decrease the
gap, and are found to restore agreement with QMC determinations in
Ref.~\cite{Gezerlis:2007fs}.  Using the numerical integration
techniques outlined in this paper, and making use of existing
higher-order calculations of the self energy~\cite{Platter:2002yr}, it
may prove feasible and worthwhile to compute the N$^3$LO corrections
to the gap.  In addition, this work motivates more precise QMC
simulations across a range of $k_F a$ to further validate the
universal corrections presented here.

In nuclear physics, the s-wave scattering lengths are very large, and
the results obtained here are only relevant at very low densities.  To
access phenomenologically interesting densities found in neutron stars
will require a more complex interaction that includes effective range
and other momentum-dependent modifications.  This can be achieved by
treating range corrections in perturbation theory, or by working in an
EFT which sums range corrections to all orders~\cite{Beane:2000fi}.  This latter method
can be implemented, for instance, by using energy-dependent potentials
as in the dimeron method~\cite{Schwenk:2005ka}, or by using
energy-independent separable
potentials~\cite{Beane:2021dab,Peng:2021pvo}.  Working with such
momentum-dependent potentials will likely extend the densities that
are accessible to perturbation theory, and will be considered
elsewhere.

\section*{Acknowledgments}
\noindent
We would like to thank Andrey Chubukov, Valentin Hirschi and Achim Schwenk for
essential conversations.  We would also like to thank Jiunn-Wei Chen, Yuki Fujimoto,
Mia Kumamoto, William Marshall and Sanjay Reddy for valuable
conversations.  We are particularly grateful to Sasha Krassovsky for
helping with the C implementation of the Monte Carlo integration.
This work was enabled, in part, by the use of advanced computational, storage and
networking infrastructure provided by the Hyak supercomputer system at
the University of Washington.  This work was supported by the Swiss
National Science Foundation (SNSF) under grant numbers 200021\_192137 and PCEFP2\_203335,
by the U.~S.~Department of Energy grant {\bf DE-FG02-97ER-41014} (UW
Nuclear Theory) and by the U.~S.~Department of Energy grant {\bf
  DE-SC0020970}, (InQubator for Quantum Simulation).

\clearpage

\appendix
\section{Superfluid gap from the BCS equations}
\label{app:scbcs}

\begin{figure*}
 \centerline{\includegraphics[width=0.3\textwidth]{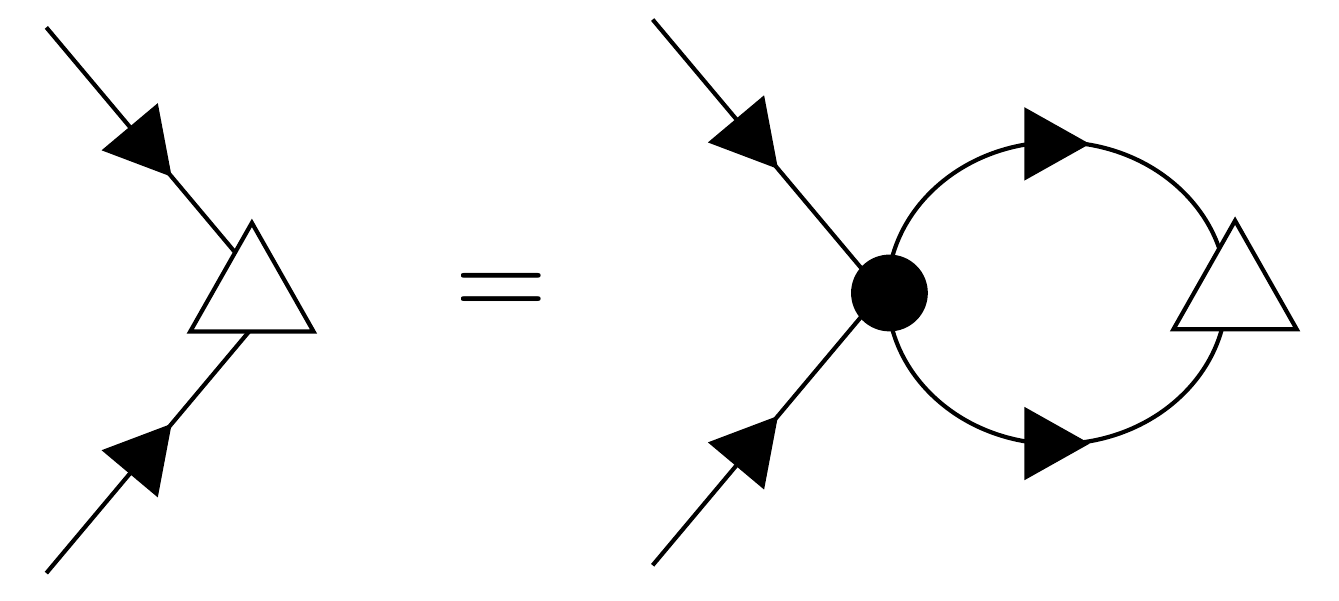}}
\caption{The BCS integral equation for the superfluid gap, represented by the empty triangle.}
\label{fig:bcvlo}
\end{figure*}

\noindent
In the presence of a Fermi surface, any attractive interaction leads
to superfluidity at zero temperature~\cite{Shankar,Polchinski:1992ed}.
The superfluid gap may be obtained from the BCS equations, which are
readily derived in the grand canonical ensemble using the Matsubara
formalism.  Supplementing the Lagrange density, Eq.~(\ref{eq:lag}),
describing the fermion interactions in free space, with a chemical
potential term, $\mu_F \psi^\dagger \psi$ and transforming to
Euclidean space allows the thermodynamic potential $\Omega(\mu_F,T)$
to be identified with the effective action.  A Hubbard-Stratonovich
transformation can then be carried out to render the four-fermion
interaction quadratic by introducing bosonic fields. Ignoring the
fluctuations in the bosonic fields, and evaluating them at their
equilibrium values, leads to the BCS mean-field approximation. In
particular, minimizing the thermodynamic potential, and taking the
zero-temperature limit, one finds the equation for the gap,
\begin{eqnarray}
\Delta({\bf k}) &=& - \int \! \frac{d^{3}{\bf  q}}{(2\pi)^{3}}\, \frac{V\left({\bf k},{\bf q}\right) \Delta({\bf q})}{2\sqrt{\left(\omega_{ q}-\mu_F\right)^2 + |\Delta({\bf q})|^2}}  \ .
  \label{eq:genbcsgapeqcont}
\end{eqnarray}
For the momentum-independent interaction considered in this work $V\left({\bf k},{\bf q}\right)=g$ . Diagrammatically, this integral equation is given in Fig.~\ref{fig:bcvlo}.
The equation for the density is finite and obtained by differentiating the thermodynamic potential with respect to $\mu_F$:
\begin{eqnarray}
\rho  & = & \frac{k_F^3}{3\pi^2} \ =\ \int \! \frac{d^3 {\bf q}}{(2\pi)^3} \Bigg\lbrack 1 \ -\ 
\frac{\left(\omega_{q}-\mu_F \right)}{\sqrt{ \left(\omega_{ q}-\mu_F \right)^2 + |\Delta({\bf q})|^2}} \Bigg\rbrack \ .
\label{eq:gapdensitygen}
\end{eqnarray}
Note that the density of the free Fermi gas is not changed by the
interaction, as an interaction that conserves particle number will
simply shift the single-particle levels, which in the ground state are still filled up to $k_F$.

Choosing the ansatz $\Delta({\bf q})=\hat\Delta_0(k_F)$, in the EFT considered here, the gap equation is:
\begin{eqnarray}
\frac{1}{g} &=& -\left(\frac{\mu}{2}\right)^{3-d} \int \! \frac{d^{d}{\bf  q}}{(2\pi)^{d}}\, \frac{1}{2\sqrt{\left(\omega_{ q}-\mu_F\right)^2 + \hat\Delta_0^2}}  \ .
  \label{eq:genbcsgapeqcontloeft}
\end{eqnarray}
It is straightforward to evaluate this linearly-divergent integral in DR with PDS~\cite{Furnstahl:2006pa,Birse:2004ha,Papenbrock:1998wb}. One finds
\begin{eqnarray}
  \frac{1}{g_R}\;=\; \frac{1}{g} + \frac{M \mu}{4\pi}&=& \frac{M}{4\pi a} \;=\; \frac{M^{3/2}}{2\sqrt{2}\pi} \mu_F^{1/2} \left( 1+ x^2 \right)^{1/4} P_{1/2}\left(-\left( 1+ x^2 \right)^{-1/2}\right)
   \label{eq:genbcsgapeqcontloeft2}
\end{eqnarray}
where $x\equiv \hat\Delta_0/\mu_F$, $P_n$ is a Legendre polynomial, and the renormalization prescription of Eq.~(\ref{eq:matchsla}) has been adopted. Finally, one obtains
\begin{eqnarray}
\frac{1}{(2 M \mu_F)^{1/2}\,a} \;=\;  \left( 1+ x^2 \right)^{1/4} P_{1/2}\left(-\left( 1+ x^2 \right)^{-1/2}\right) \ .
   \label{eq:genbcsgapeqcontloeft3}
\end{eqnarray}
As the right-hand side is negative definite in the interval $0\leq x\leq 1$, this equation has a solution only for $a<0$; that
is, for attractive interaction. The equation for the density is finite and is easily solved to give
\begin{eqnarray}
  \rho  & = & \frac{k_F^3}{3\pi^2} \ =\ -\frac{(2 M \mu_F)^{3/2}}{4\pi} \left( 1+ x^2 \right)^{1/4}\bigg\lbrack P_{1/2}\left(-\left( 1+ x^2 \right)^{-1/2}\right) \nonumber \\
  &&\qquad\qquad\qquad\qquad\qquad\qquad\ +\ \left( 1+ x^2 \right)^{1/2} P_{3/2}\left(-\left( 1+ x^2 \right)^{-1/2}\right)  \bigg\rbrack\ .
\label{eq:gapdensity2}
\end{eqnarray}
Now, given the fixed inputs $a$ and $M$, and the variable $k_F$, the
coupled equations, Eq.~(\ref{eq:genbcsgapeqcontloeft3}) and
Eq.~(\ref{eq:gapdensity2}), determine $\hat\Delta_0(k_F)$ and
$\mu_F(k_F)$.  As seen above, these results are corrected by in-medium
effects which enter as perturbations of the potential (see
Fig.~\ref{fig:GamEff}). Therefore, they rely on the potential being
weak, that is, $|k_F a|\ll 1$, which is achieved with a small
scattering length and/or a small density.

As the gap is expected to vanish for small coupling, the expressions
for the gap and for the density should admit an expansion in $x$:
\begin{eqnarray}
\frac{1}{(2 M \mu_F)^{1/2}\,a} &=&  \frac{2}{\pi}\left(2 +\log(x/8)\right) \ +\ {\mathcal O}( x^2 ) \ ,  \\
  \rho  & = & \frac{k_F^3}{3\pi^2} \ =\ \frac{(2 M \mu_F)^{3/2}}{3\pi^2} \ +\ {\mathcal O}( x^2 ) \ .
\label{eq:genbcsgapeqcontloeft3expand}
\end{eqnarray}
Therefore, neglecting ${\mathcal O}( x^2 )$ corrections, one finds from the second equation, $\mu_F=\omega_{k_F}$, and
from the first equation, the superfluid gap,
\begin{equation}
\hat\Delta_0(k_F)\ =\  \Delta_{{\rm BCS}} \ =\  \frac{8}{e^2} \omega_{k_F} \exp\left(\frac{\pi}{2 k_F a}    \right) \ .
\label{eq:singgap}
\end{equation}
As the gap is exponentially small when $|k_F a|\ll 1$, these results are accurate up to exponentially suppressed
effects at weak coupling. In particular, it is clear from Eq.~(\ref{eq:genbcsgapeqcontloeft3expand}) that the
chemical potential will experience a shift away from the Fermi energy at strong coupling.
Note that momentum-dependent effects can be estimated by replacing the scattering length
with a momentum-dependent scattering length which subsumes effective-range corrections.

\section{Including a chemical potential in the propagator}
\label{app:mu}
\noindent
The energy of particles ${\varepsilon}(k)$ at the Fermi surface is given by the chemical potential ${\varepsilon}(k_F) = \mu_F$.
In an interacting system this gets shifted away from $\omega_{k_F}$~\cite{landau1980course,Fetter}
\begin{align}
\mu_F \ = \ \omega_{k_F}\left (1 + \frac{4}{3\pi}k_F a \ + \ \frac{4(11-2\log{2})}{15\pi^2}(k_F a)^2 +\mathcal{O}[(k_F a)^3] \right ) \ ,
\end{align}
and leads to an overall shift in the denominator relative to the free propagator\footnote{Note that self-energy corrections which arise at the same
order and lead to effective mass $M^{\star}$ and wavefunction renormalization $Z$ are omitted here for simplicity.}
\begin{align}
G (k_0,{\bf k}) &=\ 
    \frac{\theta(k-k_F)}{k_0-\omega_{ k}+(\omega_{k_F}-\mu_F)+i\epsilon}
      +\frac{\theta(k_F-k)}{k_0-\omega_{ k}+(\omega_{k_F}-\mu_F)-i\epsilon} \ .
\end{align}
For the NNLO calculation of the gap, the $k_F a$ dependence of $\mu_F$ should be kept in the LO and NLO diagrams in Fig.~\ref{fig:LONLOGap} i.e. in $\Pi_{pp}(E)$ and $V^{(g^2)}$.
From Eq.~\eqref{eq:Piph} it can be seen that $\mu_F$ can be absorbed in a shift of the loop energy variable, and therefore has no effect on $V^{(g^2)}$.

The  $\mu_F$ dependence on the gap coming from $\Pi_{pp}(E)$ can be computed explicitly,
\begin{align}
\Pi_{pp}(E) \ = \ \frac{iM}{4\pi}\sqrt{2M[E + (\omega_{k_F}-\mu_F)]} \ +  \frac{M}{2 \pi^2} \bigg\lbrack \dashint_0^{k_F} &dl \, \log{\left (2M(E + (\omega_{k_F}-\mu_F))-l^2 \right)} \nonumber \\
    &- \ k_F\log{(2M(E+(\omega_{k_F}-\mu_F)) -k_F^2)}  \bigg\rbrack\ .
\label{eq:c0IBP2}
\end{align}
The pole in energy is now at $E = \mu_F + i\frac{\Delta}{2}$.
Inserting this gives
\begin{align}
\Pi_{pp}(E) \ &= \ \frac{iM}{4\pi}k_F \ + \ \frac{M}{2 \pi^2} \left [ \int_0^{k_F} dl \, \log{\left (k_F^2-l^2 \right)} \ - \ k_F\log{(i M\Delta )}\right ] \ + \ \mathcal{O}(\Delta) \nonumber \\
&=\frac{Mk_F}{2 \pi^2} \left [2(\log{2}-1) \ - \ \log{\left (\frac{\Delta}{2\omega_{k_F}}\right )}  \right ]
\ ,
\label{eq:c0IBP3}
\end{align}
in agreement with Eq.~\eqref{eq:gPipp}.
Therefore, the gap gets its ``units" from $\omega_{k_F}$, not $\mu_F$, and to NNLO it is consistent to ignore effects coming from $\mu_F-\omega_{k_F}\neq 0$ as has been done in the main text.

It is instructive to consider the more general case of a propagator with the dispersion relation left arbitrary,
\begin{align}
G (k_0,{\bf k}) &=\ 
    \frac{\theta(k-k_F)}{k_0-\varepsilon(k)+i\epsilon}
      +\frac{\theta(k_F-k)}{k_0-\varepsilon(k)-i\epsilon} \ .
\end{align}
Relevant to the ``units" of the gap is the in-medium part of $\Pi_{pp}$(E),
\begin{align}
&\dashint \frac{d^3 {\bf l}}{(2\pi)^3} \frac{\theta(k_F - l)}{E - \varepsilon(l)} \ = \ \frac{1}{2\pi^2}\dashint_0^{k_F}dl \, l^2 \frac{1}{E-\varepsilon(k_F) - (l-k_F)v_{F}  + {\mathcal O}[(l-k_F)^2]}\nonumber \\
&= \ -\frac{1}{4\pi^2 v_F^3}\left [k_F v_F(2E-2\varepsilon(k_F) +3 k_F v_F)\ + \ 2 (E-\varepsilon(k_F)+k_F v_F)^2\log{\left (\frac{\varepsilon(k_F)-E}{E-\varepsilon(k_F)+k_F v_F}\right )} \right ] \nonumber \\
&\stackrel{E \to \varepsilon(k_F) + i\Delta/2}{=} \ -\frac{k_F^2}{4\pi^2 v_F}\left [3 \ + \ 2\log{\left (\frac{-i\Delta}{2k_F v_F} \right )} \right ] + {\mathcal O}(\Delta)\ ,
\end{align}
where $v_F = \frac{d}{dk}\varepsilon(k)\big \vert_{k_F}$ is the velocity of quasi-particles at the Fermi surface.
The third line has been evaluated at the pole $E = \varepsilon(k_F) + i\Delta/2$.
This shows that the ``units" of the gap come from $v_F k_F$, and are therefore unaffected by a constant energy shift to $\varepsilon(k)$, i.e. by the chemical potential $\mu_F$.

\section{Details of the NNLO Calculation}
\label{app:cdNNLO}
\begin{figure}
    \centering
    \includegraphics[width=\textwidth]{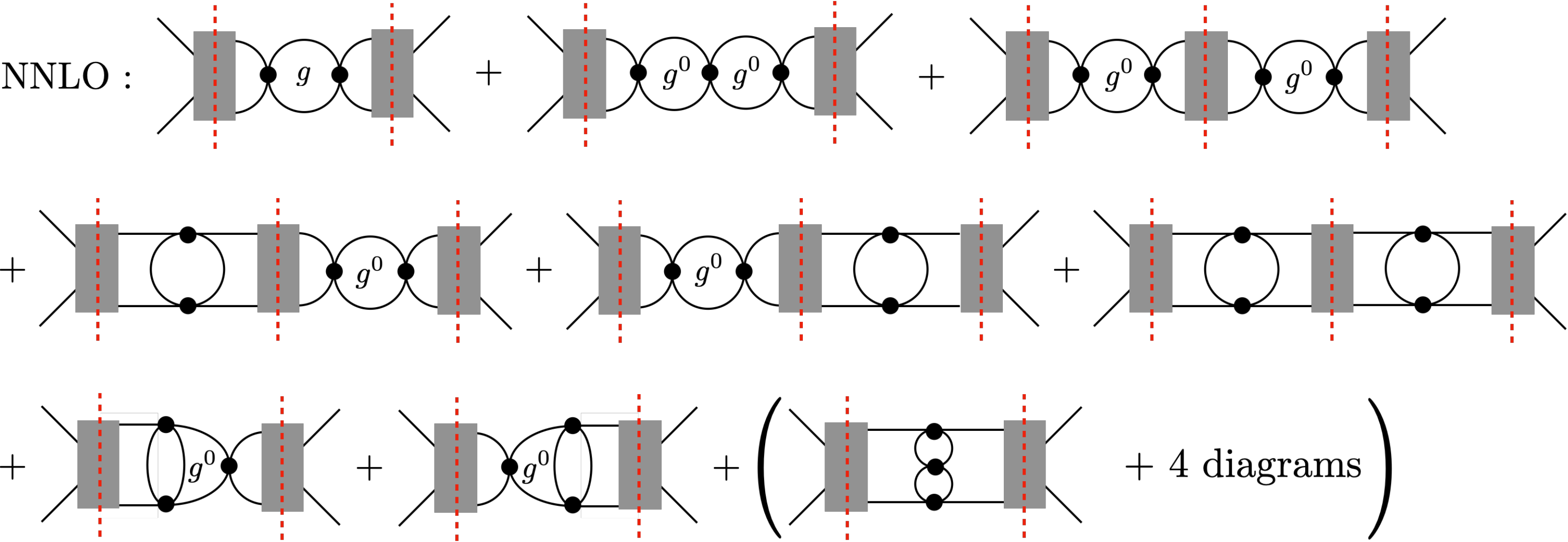}
    \caption{New diagrams contributing to the vertex function for the NNLO determination of the gap.}
    \label{fig:NNLOdiag}
\end{figure}

\noindent
The diagrams that contribute to the NNLO gap are shown in Fig.~\ref{fig:NNLOGap}.
For completeness, the new diagrams at this order are explicitly given in Fig.~\ref{fig:NNLOdiag}.
It is straightforward to evaluate diagram 3a):
\begin{align}
\Gamma_{3a} \ = \ -\frac{g}{1+g \Pi_{pp}^{(g^{-1})}} \left [1- \frac{g \Pi_{pp}^{(g^0)} + g \Pi_{pp}^{(g)}}{1+g \Pi_{pp}^{(g^{-1})}} + \left (\frac{g \Pi_{pp}^{(g^0)}}{1+g \Pi_{pp}^{(g^{-1})}}\right )^2\right ] \ ,
\label{eq:Gam3a}
\end{align} 
diagram 3b):
\begin{align}
    \Gamma_{3b} \ = \ \frac{-V^{(g^2)}}{\left ( 1+g\Pi_{pp}^{(g^{-1})}\right )^2} \left ( 1- 2\frac{g \Pi_{pp}^{(g^0)} }{1+g \Pi_{pp}^{(g^{-1})}}  \right ) \ ,
\end{align}
and diagram 3c):
\begin{align}
    \Gamma_{3c} \ = \ \frac{\left (V^{(g^2)}\right )^2 \, \Pi_{pp}^{(g^{-1})}}{\left ( 1+g\Pi_{pp}^{(g^{-1})}\right )^3} \ = \ \frac{-\left (V^{(g^2)}\right )^2 }{g \left ( 1+g\Pi_{pp}^{(g^{-1})}\right )^3} \ .
\end{align}
In the second equality for $\Gamma_{3c}$ the LO result $\Pi_{pp}^{(g^{-1})} = -1/g$ has been used.
The evaluation of diagram 3d) follows similarly to 2b) and gives
\begin{align}
\Gamma_{3d} \ = \ \frac{-V^{(g^3)}}{\left ( 1+g\Pi_{pp}^{(g^{-1})}\right )^2}  \ ,
\label{eq:Gam3d}
\end{align}
where $V^{(g^3)}$ is determined from the diagrams that contribute to the ${\mathcal O}(g^3)$ component of the 2PI potential, shown in Fig.~\ref{fig:NNLOV}.
Diagram 3e) evaluates to
\begin{align}
\Gamma_{3e} \ = \ 2\frac{g\, {\cal I}\left [\Pi^{(g^0)}_{pp}\,V^{(g^2)}\right ]}{\left ( 1+g\Pi_{pp}^{(g^{-1})}\right )^2}  \ ,
\label{eq:Gam3e}
\end{align}
where ${\cal I}\left [\Pi^{(g^0)}_{pp}\,V^{(g^2)}\right ]$ is the integral of the p-p bubble with ${\cal O}(g^{-1})$ piece subtracted, combined with the ${\cal O}(g^2)$ component of the 2PI potential. See App.~\ref{app:Vl}.

The Feynman diagrams in $V^{(g^3)}$ and ${\cal I}\left [\Pi^{(g^0)}_{pp}\,V^{(g^2)}\right ]$ are evaluated using a numerical
technique originally developed for relativistic quantum field theory.
This technique is based on deriving the three-dimensional
representation of a Feynman diagram by integrating the energy
components of loop integrals
analytically~\cite{Catani_2008,Bierenbaum_2010,Runkel_2019,Capatti:2019ypt,Capatti:2022mly,Aguilera_Verdugo_2021,Sborlini_2021},
performing the singularity analysis and regularization in the
integration space of the remaining spatial
components~\cite{Capatti:2019edf,Kermanschah:2021wbk,Capatti:2022tit},
and finally numerically performing the regularized integral by direct
Monte-Carlo integration (see~\cite{Capatti:2023omc} for a
comprehensive review of the method
and~\cite{AH:2023kor,Navarrete:2024zgz} for recent applications).  The
results presented below used the \texttt{VEGAS} Python
package~\cite{Lepage:2020tgj} with {\tt nitn=20} and {\tt
  neval=$1\times 10^{11}$}, except for the evaluation of $\Pi^{(a)}_{ppph}$ which used {\tt
  neval=$5\times10^{10}$}.

\subsection{Computing \texorpdfstring{$V^{(g^3)}$}{}}
\label{app:VNNLO}
\begin{figure}
    \centering
    \includegraphics[width=0.5\textwidth]{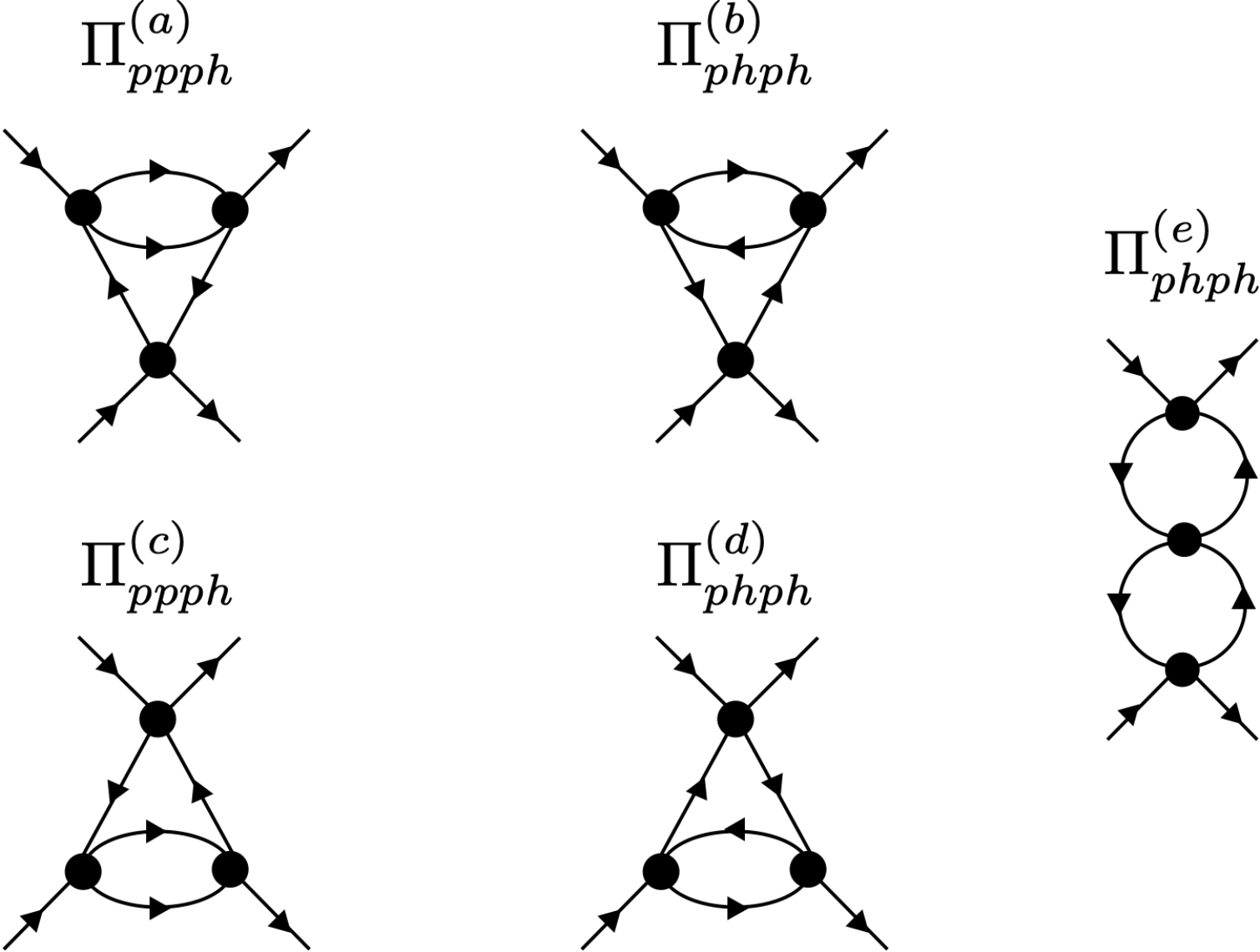}
    \caption{Diagrams contributing to the 2PI potential at ${\mathcal O}(g^3)$.}
    \label{fig:NNLOV}
\end{figure}

\begin{figure}
    \centering
    \includegraphics[width=0.8\textwidth]{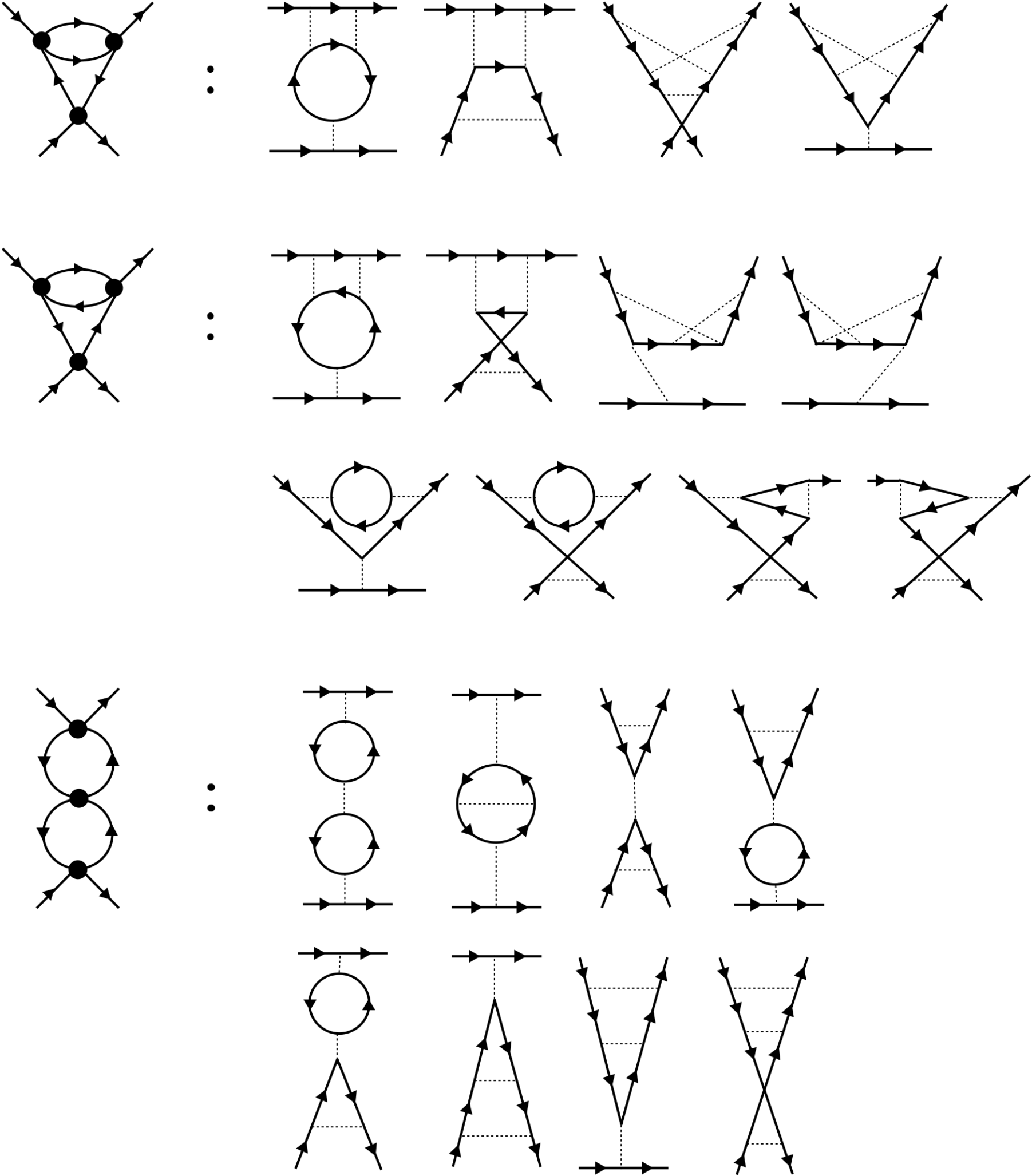}
    \caption{Spin contractions for $V^{(g^3)}$.}
    \label{fig:NNLODOT}
\end{figure}

\noindent
Determining the NNLO gap requires computing $V^{(g^3)}$, the
${\mathcal O}(g^3)$ piece of the 2PI s-wave potential evaluated for
external legs on-shell and at the Fermi surface.  The diagrams that
contribute are shown in Fig.~\ref{fig:NNLOV}, and their evaluation can
be simplified by noting that $\Pi_{ppph}^{(c)}$ and $\Pi_{phph}^{(d)}$
are related to $\Pi_{ppph}^{(a)}$ and $\Pi_{phph}^{(b)}$ by ${\bf k}
\to -{\bf k}$, and ${\bf k}' \to -{\bf k}'$.
Applying the Feynman rules yields:
\begin{eqnarray}
\Pi_{ppph}^{(a)}({\bf k},{\bf k}') & = & \int \!  \frac{d^4 l}{(2\pi)^4} \int \! \frac{d^4 l'}{(2\pi)^4} G_0(E+l_0,{\bf l}) G_0(l_0'-l_0,{\bf l}'-{\bf l}) G_0(l_0',{\bf l}'-{\bf k}) G_0(l_0',{\bf l}'-{\bf k}') \ , \\
\Pi_{phph}^{(b)}({\bf k},{\bf k}') & = & \int \!  \frac{d^4 l}{(2\pi)^4} \int \! \frac{d^4 l'}{(2\pi)^4} G_0(E+l_0-l_0',{\bf l}-{\bf l}'+{\bf k}'+{\bf k}) G_0(l_0,{\bf l}) \nonumber \\
&&\mkern300mu \times G_0(l_0',{\bf l}'-{\bf k}')G_0(l_0',{\bf l}'-{\bf k}) \ ,\\
\Pi_{phph}^{(e)}({\bf k},{\bf k}') &= & \int \! \frac{d^4 l}{(2\pi)^4} \int \! \frac{d^4 l'}{(2\pi)^4} G_0(l_0,{\bf l}+{\bf k}-{\bf k}') G_0(l_0,{\bf l}) G_0(l_0',{\bf l}'+{\bf k}-{\bf k}') G_0(l_0',{\bf l}') \nonumber \\
&=&  -\left [\Pi_{ph}({\bf k},{\bf k}';0) \right ]^2\ .
\end{eqnarray}
The functional dependence on $k_0=k_0'$ has been omitted as both energies are set to $\omega_{k_F}$.
Putting these together, the ${\cal O}(g^3)$ contribution to the full 2PI potential is
\begin{align}
&V^{(g^3)}_{\alpha\beta,\gamma\delta}({\bf k},{\bf k}') \ = \nonumber \\
&g^3\bigg \{ \delta_{\alpha \delta}\delta_{\beta\gamma} \left [\Pi_{ppph}^{(a)}({\bf k},{\bf k}') +  \Pi_{ppph}^{(a)}(-{\bf k},-{\bf k}') \right ] \ - \ \delta_{\alpha \gamma}\delta_{\beta\delta} \left [\Pi_{ppph}^{(a)}({\bf k},-{\bf k}') + \Pi_{ppph}^{(a)}(-{\bf k},{\bf k}') \right ] \nonumber \\
& + \ \delta_{\alpha \gamma}\delta_{\beta\delta} \left [\Pi_{phph}^{(b)}({\bf k},{\bf k}') +  \Pi_{phph}^{(b)}(-{\bf k},-{\bf k}') \right ] \ - \ \delta_{\alpha \delta}\delta_{\beta\gamma} \left [\Pi_{phph}^{(b)}({\bf k},-{\bf k}') + \Pi_{phph}^{(b)}(-{\bf k},{\bf k}') \right ]
\nonumber \\
& + \left ( \delta_{\alpha \gamma}\delta_{\beta\delta} - \delta_{\alpha \delta}\delta_{\beta\gamma} \right )\left ([\Pi_{ph}({\bf k},{\bf k}';0) ]^2 + [\Pi_{ph}({\bf k},-{\bf k}';0) ]^2 \right ) \bigg \} \nonumber \\
&=g^3\bigg \{ 2\delta_{\alpha \delta}\delta_{\beta\gamma} \Pi_{ppph}^{(a)}({\bf k},{\bf k}')   -  2\delta_{\alpha \gamma}\delta_{\beta\delta} \Pi_{ppph}^{(a)}({\bf k},-{\bf k}')  + 2\delta_{\alpha \gamma}\delta_{\beta\delta} \Pi_{phph}^{(b)}({\bf k},{\bf k}')  -  2\delta_{\alpha \delta}\delta_{\beta\gamma} \Pi_{phph}^{(b)}({\bf k},-{\bf k}')
\nonumber \\
& + \left ( \delta_{\alpha \gamma}\delta_{\beta\delta} - \delta_{\alpha \delta}\delta_{\beta\gamma} \right )\left ([\Pi_{ph}({\bf k},{\bf k}';0) ]^2 + [\Pi_{ph}({\bf k},-{\bf k}';0) ]^2 \right ) \bigg \} \nonumber \\
& \ \stackrel{\text{s-wave}}{=}2g^3\left ( \delta_{\alpha \gamma}\delta_{\beta\delta} - \delta_{\alpha \delta}\delta_{\beta\gamma} \right )\left ([\Pi_{ph}({\bf k},{\bf k}';0) ]^2 \ - \ \Pi_{ppph}^{(a)}({\bf k},{\bf k}')\ + \ \Pi_{phph}^{(b)}({\bf k},{\bf k}')\right ) \ ,
\label{eq:VNNLO}
\end{align}
where the third equality has used the fact that the s-wave projection is invariant under ${\bf k}'\to-{\bf k}'$.
The spin factors are determined from the dotted-line decomposition shown in Fig.~\ref{fig:NNLODOT}.

To determine $V^{(g^3)}$ the potential has to be projected onto the s-wave. 
For numerical evaluation it is convenient to do this by fixing ${\bf k} = k_F \hat{z}$ and integrating over ${\bf k}'$ in the x-z plane i.e. ${\bf k}' = k_F(\sin{\theta},0,\cos{\theta})^T$.
For example,
\begin{align}
V^{g^3} \ &= \ g^3\int_0^{\pi}\! d\theta \sin{\theta}\left [\Pi(k_F \hat{z},k_F(\sin{\theta},0,\cos{\theta})^T)  \right ]\ .
\label{eq:MCsWave}
\end{align}
The final result is
\begin{align}
V^{(g^3)} \ &= g^3 \int_0^\pi \! d\theta \, \sin{\theta} \left ( -\Pi^{(a)}_{ppph}({\bf k},{\bf k}') \ + \ \Pi^{(b)}_{phph}({\bf k},{\bf k}') \ + \ \Pi_{ph}({\bf k},{\bf k}')^2\right) \nonumber \\
&= g \frac{g^2 M^2 k_F^2}{4\pi^4}\left (0.972041(15)\ - \ 0.0418477(64) \ + \ \frac{1}{4} \left [1 + \frac{7}{2}\zeta(3) \right ] \right ) \nonumber \\
&=  2.231993(16)\, g \, \lambda^2 \ .
\label{eq:l0g3Project}
\end{align}
The evaluation of the individual terms is detailed below. 

First, consider evaluating $\Pi_{ppph}^{(a)}({\bf k}, {\bf k}')$.  It
is convenient to define ${\bf q_1 } = {\bf l}' -{\bf k}'$, ${\bf q_2 }
= {\bf l}'- {\bf k}$ and ${\bf q_3 } = {\bf l}' - {\bf l}$.  Performing 
the $l_0$ and $l_0'$ integrals one finds
\begin{align}
&\Pi^{(a)}_{ppph}({\bf k},{\bf k}') \ = \ \int \! \frac{d^4l'}{(2\pi)^4}\int\! \frac{d^4l}{(2\pi)^4}G_0(l_0',{\bf q}_1)G_0(l_0',{\bf q}_2)G_0(E+l_0,{\bf l})G_0(l_0'-l_0,{\bf q}_3) \nonumber \\[4pt]
&=\int\! \frac{d^3 {\bf l}'}{(2\pi)^3}\int\! \frac{d^3 {\bf l}}{(2\pi)^3} \Bigg \{
\frac{\theta(k_F-q_3)-\theta(l-k_F) }{\omega_{q_1}-\omega_{q_2}}\left (\frac{\theta(k_F-q_2)}{\omega_{q_2}+E-\omega_l-\omega_{q_3}}-\frac{\theta(k_F-q_1)}{\omega_{q_1}+E-\omega_l-\omega_{q_3}} \right )\nonumber \\
&-\theta(k_F-l)\theta(k_F-q_3)\left [\frac{1}{(-E+\omega_l+\omega_{q_3}-\omega_{q_1})(-E+\omega_l+\omega_{q_3}-\omega_{q_2})}\right ] \Bigg \} \ .
\end{align}
In the second line, the $(i\epsilon)$s can be safely dropped because the
integrand does not have non-integrable singularities (and therefore
there is no imaginary component).  This is explicitly verified at a
mathematical level by a simple power-counting procedure, and at a
practical level by performing the numerical integration, which gives a
finite result with no need for a principal value prescription.  The
terms with $\theta(l-k_F)$ are UV divergent for $l\to\infty$, and can
be subtracted and evaluated analytically in the PDS scheme.  To
preserve the IR structure only the leading UV divergences are
subtracted,
\begin{align}
&\int\! \frac{d^3 {\bf l}'}{(2\pi)^3}\int\! \frac{d^3 {\bf l}}{(2\pi)^3} 
\frac{-\theta(l-k_F) }{\omega_{q_1}-\omega_{q_2}}\left (\frac{\theta(k_F-q_2)}{\omega_{q_2}+E-\omega_l-\omega_{q_3}}-\frac{\theta(k_F-q_1)}{\omega_{q_1}+E-\omega_l-\omega_{q_3}} \right )\nonumber \\
&=\int\! \frac{d^3 {\bf l}'}{(2\pi)^3}\int\! \frac{d^3 {\bf l}}{(2\pi)^3} \frac{-\theta(l-k_F)}{2\omega_l}\left (1+\frac{{\bf l}\cdot {\bf l}' }{l^2} + \mathcal{O}(l^{-2})\right )\left [\frac{\theta(k_F-q_1)-\theta(k_F-q_2)}{\omega_{q_1}-\omega_{q_2}} \right ]\nonumber \\
&=\frac{M \mu}{4\pi}\Pi_{ph}(q) - \frac{M k_F}{2 \pi^2}\Pi_{ph}(q) \ .
\end{align}
Notice that the ${\bf l}\cdot {\bf l}'$ term is only zero at the integrated level, and therefore needs to be kept so that the integrand is UV finite. 
Explicitly,
\begin{align}
  &\Pi^{(a)}_{ppph}({\bf k},{\bf k}') = \left (\frac{M \mu}{4\pi} - \frac{M k_F}{2 \pi^2}\right )\Pi_{ph}(q)\nonumber\\
  &+  \int\! \frac{d^3 {\bf l}'}{(2\pi)^3}\int\! \frac{d^3 {\bf l}}{(2\pi)^3} \Bigg \{
\frac{\theta(k_F-q_3)-\theta(l-k_F) }{\omega_{q_1}-\omega_{q_2}}\left (\frac{\theta(k_F-q_2)}{\omega_{q_2}+E-\omega_l-\omega_{q_3}}-\frac{\theta(k_F-q_1)}{\omega_{q_1}+E-\omega_l-\omega_{q_3}} \right )\nonumber \\
&-\theta(k_F-l)\theta(k_F-q_3)\left [\frac{1}{(-E+\omega_l+\omega_{q_3}-\omega_{q_1})(-E+\omega_l+\omega_{q_3}-\omega_{q_2})}\right ]\nonumber\\
&+ \frac{\theta(l-k_F)}{2\omega_l}\left (1+\frac{{\bf l}\cdot {\bf l}' }{l^2}\right )\left [\frac{\theta(k_F-q_1)-\theta(k_F-q_2)}{\omega_{q_1}-\omega_{q_2}} \right ]\Bigg \} \ ,
\end{align}
where the integral in curly braces is evaluated with Monte Carlo integration.

To verify renormalizability, it can be shown that all $\mu$-dependence cancels at this order.
The $\mu$-dependence in $V^{(g^3)}$ coming from $\Pi_{pphh}^{(a)}$ is,
\begin{align}
&V^{(g^3)}_{\alpha,\beta,\gamma\delta}({\bf k},{\bf k}';\omega)[\mu] \ = \ 2g(\mu)^3 \frac{M \mu}{4\pi} \left (\Pi_{ph}({\bf k},{\bf k}';\omega)\delta_{\alpha \delta}\delta_{\beta\gamma} \ - \ \Pi_{ph}({\bf k},-{\bf k}';\omega)\delta_{\alpha \gamma}\delta_{\beta\delta} \right ) \nonumber \\
&= \left ( \frac{4\pi a}{M}\right )^2 \left (2a\mu \ + \ {\mathcal O}[(a\mu)^2]\right ) \left (\Pi_{ph}({\bf k},{\bf k}';\omega)\delta_{\alpha \delta}\delta_{\beta\gamma} \ - \ \Pi_{ph}({\bf k},-{\bf k}';\omega)\delta_{\alpha \gamma}\delta_{\beta\delta} \right ) \ ,
\end{align}
where the PDS coupling has been expanded in powers of $(a\mu)$ using Eq.~(\ref{eq:gmu}).
This $\mu$-dependence must cancel in physical quantities, for example in the NNLO gap equation, Eq.~(\ref{eq:NNLOGapEqn}).
Only the sum $\left [ V^{(g^2)} + V^{(g^3)} \right ]$ enters and the $\mu$-dependence of $V^{(g^2)}$
\begin{align}
&V^{(g^2)}_{\alpha \beta, \gamma \delta}({\bf k}, {\bf k}';\omega)[\mu] \ = \  g(\mu)^2\left (\Pi_{ph}({\bf k}, -{\bf k}';\omega) \delta_{\alpha \gamma} \delta_{\beta \delta}-\Pi_{ph}({\bf k}, {\bf k}';\omega) \delta_{\alpha \delta} \delta_{\beta \gamma} \right ) \nonumber \\
& = \ \left (\frac{4 \pi a}{M}\right )^2\left (-1  \ - \ 2a \mu  \ + \ {\mathcal O}([a\mu]^2)\right )\left (\Pi_{ph}({\bf k}, {\bf k}';\omega) \delta_{\alpha \delta} \delta_{\beta \gamma}-\Pi_{ph}({\bf k}, -{\bf k}';\omega) \delta_{\alpha \gamma} \delta_{\beta \delta} \right ) \ ,
\end{align}
removes all $\mu$-dependence from the NNLO gap equation.

Next is the evaluation of $\Pi^{(b)}_{phph}({\bf k},{\bf k}')$. 
Defining ${\bf q_1 } = {\bf l}' -{\bf k}'$, ${\bf q_2 } = {\bf l}'- {\bf k}$ and ${\bf q_3 } = {\bf l} - {\bf l}' + {\bf k} + {\bf k}'$, and performing the $l_0$ and $l_0'$ integrals gives,
\begin{align}
&\Pi^{(b)}_{phph}({\bf k},{\bf k}') \ = \ \int \! \frac{d^4l'}{(2\pi)^4}\int\! \frac{d^4l}{(2\pi)^4}G_0(l_0',{\bf q}_1)G_0(l_0',{\bf q}_2)G_0(l_0,{\bf l})G_0(E+l_0-l_0',{\bf q}_3) \nonumber \\[4pt]
&=\int\! \frac{d^3 {\bf l}'}{(2\pi)^3}\int\! \frac{d^3 {\bf l}}{(2\pi)^3} \Bigg \{\frac{\theta(k_F-l)-\theta(k_F-q_3)}{\omega_{q_1}-\omega_{q_2}}\left [\frac{\theta(q_1-k_F)}{-\omega_{q_1}+E+\omega_l-\omega_{q_3}} \ - \ \frac{\theta(q_2-k_F)}{-\omega_{q_2}+E+\omega_l-\omega_{q_3}}\right ]\nonumber \\
 &+ \ \theta(l-k_F)\theta(k_F-q_3)\left [\frac{1}{(E+\omega_l-\omega_{q_3}-\omega_{q_1})(E+\omega_l-\omega_{q_3}-\omega_{q_2})}\right ] \Bigg \} \ .
\end{align}
This expression has no non-integrable singularities, and the contribution to the s-wave potential can be directly computed using Eq.~(\ref{eq:MCsWave}).

As a check, the potential can be projected onto the p-wave, and compared to Eq.~(18) in Ref.~\cite{Efremov_2000}. The p-wave potential is 
\begin{align}
V^{g^3}_{\ell = 1} \ &= \ g^3\int_0^{\pi} \! d\theta \sin{\theta}\cos{\theta}\left [\Pi^{(a)}_{ppph}({\bf k},{\bf k}') \ + \ \Pi^{(b)}_{phph}({\bf k},{\bf k}')\right ] \nonumber \\
& =-0.3313(30)g\lambda^2
\end{align}
in agreement with $V^{g^3}_{\ell = 1} =-0.33g\lambda^2$ quoted in  Ref.~\cite{Efremov_2000}.
Note that $\Pi^{(e)}_{phph}=-
(\Pi_{ph})^2$ does not contribute for odd partial waves, and that ${\bf k}' \to -{\bf k}'$ causes a change of sign (and hence no minus in front of $\Pi^{(a)}$).

\subsection{Computing \texorpdfstring{$\Gamma_{3e}$}{}}
\label{app:Vl}

\noindent
Determining the NNLO gap requires computing diagram 3e) in Fig.~\ref{fig:NNLOGap},
\begin{align}
\Gamma_{3e} \ &= \ \frac{2g^3}{\left [1+g \Pi_{pp}^{(g^{-1})} \right ]^2} \int \! \frac{d^4 l}{(2\pi)^4}\int \! \frac{d^4 l'}{(2\pi)^4} G_0({\bf l},l_0+E)G_0(-{\bf l},-l_0+E)
G(l_0',{\bf l}')G(l_0'-l_0,q_3) \nonumber \\
&\equiv  \frac{2 g}{\left [1+g \Pi_{pp}^{(g^{-1})} \right ]^2} \, {\cal I}\left [\Pi^{(g^0)}_{pp}\,V^{(g^2)}\right ],
\end{align}
where ${\bf q}_3 = {\bf l}'+{\bf k}' - {\bf l}$ with ${\bf k}'$ the outgoing momentum. 
Since, the particle-particle bubbles do not mix partial waves, the final result will not depend on ${\bf k}'$, but intermediate results will.
Performing the two energy integrals gives,
\begin{align}
&{\cal I}\left [\Pi^{(g^0)}_{pp}\,V^{(g^2)} \right ] \ = \ g^2\int \! \frac{d^3 {\bf l}}{(2\pi)^3}\int \! \frac{d^3 {\bf l}'}{(2\pi)^3}\Bigg \{ \frac{\theta(l-k_F)\theta(l'-k_F)\theta(k_F-q_3)}{(E+\omega_{l'}-\omega_{q_3}-\omega_l+i\epsilon)(E+\omega_{q_3}-\omega_{l'}-\omega_l+i\epsilon)} \nonumber \\ &+\frac{\theta(k_F-l)\theta(l'-k_F)\theta(k_F-q_3)}{(E+\omega_{l'}-\omega_{q_3}-\omega_l-i\epsilon)(E+\omega_{q_3}-\omega_{l'}-\omega_l-i\epsilon)} 
+\frac{\theta(l-k_F)\left (\theta(k_F-l')-\theta(k_F-q_3) \right )}{(2E-2\omega_l+i\epsilon)(\omega_{l'}+E-\omega_l-\omega_{q_3}+i\epsilon)} 
 \nonumber \\
&-\frac{\theta(k_F-l)\left (\theta(k_F-l')-\theta(k_F-q_3) \right )}{(2E-2\omega_l-i\epsilon)(-E+\omega_{l'}
+\omega_l-\omega_{q_3}+i\epsilon)} \Bigg \}\nonumber \\
&= g^2\int \! \frac{d^3 {\bf l}}{(2\pi)^3}\int \! \frac{d^3 {\bf l}'}{(2\pi)^3}\Bigg \{\frac{\theta(l'-k_F)\theta(k_F-q_3)}{(E+\omega_{l'}-\omega_{q_3}-\omega_l)(E+\omega_{q_3}-\omega_{l'}-\omega_l)} \nonumber \\
&+\frac{\theta(l-k_F)\left (\theta(k_F-l')-\theta(k_F-q_3) \right )}{(2E-2\omega_l+i\epsilon)(\omega_{l'}+E-\omega_l-\omega_{q_3})} -\frac{\theta(k_F-l)\left (\theta(k_F-l')-\theta(k_F-q_3) \right )}{(2E-2\omega_l-i\epsilon)(-E+\omega_{l'}
+\omega_l-\omega_{q_3})} \Bigg \}
\end{align}
where in the second equality the $(i\epsilon)$s which regulate integrable singularities have been dropped.
The only non-integrable singularity is at $2E = 2\omega_l$, corresponding to the BCS singularity in the $\Pi_{pp}$ sub-loop. 
This can be regulated by subtracting and adding back a counter-term where $\omega_l = E$ in all non-singular quantities,
\begin{align}
& \int \! \frac{d^3 {\bf l}}{(2\pi)^3}\int \! \frac{d^3 {\bf l}'}{(2\pi)^3}\frac{\left (\theta(k_F-l')-\theta(k_F-\bar{q}_3) \right )\left (\theta(\Lambda-l)\theta(l-k_F)- \theta(k_F-l)\right )}{(\omega_{l'}-\omega_{\bar{q}_3})(2E-2\omega_l)}\nonumber \\
&=\int \!\frac{d^3 {\bf l}'}{(2\pi)^3}\int \! \frac{d\Omega_l}{(2\pi)^3}\frac{\left (\theta(k_F-l')-\theta(k_F-\bar{q}_3) \right )}{(\omega_{l'}-\omega_{\bar{q}_3})}\left [\int_0^{\Lambda}\! dl \frac{l^2}{2E-2\omega_l}-2\int_0^{k_F}\! dl\frac{l^2}{2E-2\omega_l}  \right ] \nonumber \\
&=-\frac{k_F M}{4\pi^2} \int_0^2 \! dq \Pi_{ph}(q)\left [-\frac{\Lambda}{k_F} + \coth^{-1}\left (\frac{\Lambda}{k_F}\right ) -2(\log{2}-1)+\log \left ( \frac{\Delta}{2\omega_{k_F}}\right ) \right ] \nonumber \\
&=-\frac{M^2 k_F^2}{12\pi^4}(1+2\log{2})\left [-\frac{1}{3}(1+2\log{2}) -\frac{\Lambda}{k_F} + \coth^{-1}\left ( \frac{\Lambda}{k_F}\right ) \right ]
\end{align}
where $\overline{q}_3 = {\bf l}' +{\bf k}' -k_F {\bf l}/l$ is $q_3$ with $l =k_F$. 
This decouples the singular integration over $l$ in the two loops. Taking $q_3 \to \bar{q}_3$ changes the UV behavior of the $\int \! d^3 l$ integration, and has been regulated with the $\theta(\Lambda -l)$.
As this quantity is subtracted and added back in, the final result will not depend on $\Lambda$.
In the fourth line, the ${\cal O}(g^0)$ expression for $\log{\left ({\Delta}/(2 \omega_{k_F})\right )}$ has been substituted since the ${\cal O}(g^{-1})$ piece has already been counted in $\Gamma_{3b}$.
Explicitly, 
\begin{align}
  &{\cal I}\left [\Pi^{(g^0)}_{pp}\,V^{(g^2)} \right ] \ = \ g^2\int \! \frac{d^3 {\bf l}}{(2\pi)^3}\int \! \frac{d^3 {\bf l}'}{(2\pi)^3}\Bigg \{\frac{\theta(l'-k_F)\theta(k_F-q_3)}{(E+\omega_{l'}-\omega_{q_3}-\omega_l)(E+\omega_{q_3}-\omega_{l'}-\omega_l)}\nonumber \\
  &+\frac{\theta(l-k_F)\left (\theta(k_F-l')-\theta(k_F-q_3) \right )}{(2E-2\omega_l)(\omega_{l'}+E-\omega_l-\omega_{q_3})}
-\frac{\theta(k_F-l)\left (\theta(k_F-l')-\theta(k_F-q_3) \right )}{(2E-2\omega_l)(-E+\omega_{l'}
  +\omega_l-\omega_{q_3})}\nonumber \\ 
&- \frac{\left (\theta(k_F-l')-\theta(k_F-\bar{q}_3) \right )\left (\theta(\Lambda-l)\theta(l-k_F)- \theta(k_F-l)\right )}{(\omega_{l'}-\omega_{\bar{q}_3})(2E-2\omega_l)}\Bigg \} \nonumber \\
&- \frac{M^2 k_F^2}{12\pi^4}(1+2\log{2})\left [-\frac{1}{3}(1+2\log{2}) -\frac{\Lambda}{k_F} + \coth^{-1}\left ( \frac{\Lambda}{k_F}\right ) \right ]
\end{align}
where the expression in curly braces is evaluated with Monte Carlo.
The final result does not depend on ${\bf k}'$ or $\Lambda$, and is found to be
\begin{equation}
{\cal I}\left [\Pi^{(g^0)}_{pp}\,V^{(g^2)} \right ] \ = \  0.917746(98) \, \lambda^2 \ .
\end{equation}

\bibliography{bibi}

\end{document}